\documentclass[12pt,english]{article}
\usepackage[T1]{fontenc}
\usepackage[latin9]{inputenc}
\usepackage[a4paper]{geometry}
\usepackage{amsmath}
\usepackage{amssymb}
\usepackage{graphicx}
\usepackage{esint}

\makeatletter
\numberwithin{equation}{section}

\makeatother

\usepackage{babel}
\begin{document}

\title{A new class of fuzzy spaces with classical limit}

\author{Andreas Sykora\thanks{syko@gelbes-sofa.de}}

\date{September 26, 2017}
\maketitle
\begin{abstract}
We present a new type of matrix regularization, which is based on
matrix-valued functions defined on a cylinder. If non-commutative
coordinates of a fuzzy space are defined by a regularization of such
functions, we show that a classical limit for the fuzzy spaces exists,
when the matrix-valued functions nearly commute. In this case, the
classical limit of the fuzzy space is a manifold, which is composed
of coordinate patches that are defined by the diagonal entries of
the diagonalized matrix-valued functions. As applications, an interpolation
for direct sums of fuzzy spaces is described and a fuzzy string vertex
with classical limit, i.e. a surface interconnecting three circles,
is constructed.

\newpage{}
\end{abstract}
\tableofcontents{}

\newpage{}

\section{Introduction}

Non-commutative geometry is a promising candidate for describing quantized
space-times, at least for providing effective actions, which go beyond
the classical regime of general relativity \cite{Doplicher:1995}.
In non-commutative geometry, the commutative algebra of functions
on a manifold is replaced with a non-commutative algebra having extra
structures that describe the geometry of the space. One approach for
this are ``spectral triples'', which encode the geometry of the
non-commutative space in a functional analytic way \cite{Connes:1994}.

We follow here a more physical approach, which is motivated by string
theory and matrix models \cite{Ishibashi:1997,Banks:1997,Steinacker:2011}.
In this approach, which we call ``fuzzy spaces'', the non-commutative
geometry is defined analogously to an embedding of a classical manifold
in a target space. In particular, a classical manifold can be defined
via its coordinate functions, which embed it in a higher dimensional
space, which is usually $\mathbb{R}^{n}$. Other features of the geometry,
such as a metric, then can be defined via a pull-back from the embedding
space. For fuzzy spaces, the coordinate functions are replaced with
coordinate matrices or more general with operators on a Hilbert space
and it is then tried to derive objects such as a non-commutative metric,
curvature etc., analogously to the corresponding objects in a classical
geometry \cite{Madore:1999}. 

One way to define a fuzzy space is via algebraic relations of the
non-commutative coordinates. Representations of the generated algebra
then result in matrices for the non-commutative coordinates, which
have the demanded properties. Such an approach is feasible for highly
symmetric spaces, such as the fuzzy sphere \cite{Madore:1991} or
the fuzzy torus \cite{Steinacker:2011}, but becomes more and more
complicated, in particular when the algebra relations or commutators
become non-polynomial functions.

Another approach for defining fuzzy spaces is matrix regularization
based on Fourier decomposition (see \cite{Shimada:2003}, for example).
The algebra of functions on a manifold is decomposed into Fourier
series and the Fourier series are truncated at a specific order. The
Hilbert space basis of the Fourier series can be mapped to a basis
of matrices and in such a way, every truncated function can be mapped
to a matrix. This is the approach we will extend in the following
chapters.

In chapter 2 we investigate the spaces introduced in \cite{Shimada:2003},
which we call ``immersed cylinder fuzzy spaces'', since these spaces
are matrix regularizations of a cylinder immersed in $\mathbb{R}^{n}$.
When it is possible to define the manifold to be regularized via only
one coordinate patch in the form of a cylinder, i.e. a circle times
an interval, the coordinate functions, which immerse the cylinder,
can be decomposed into their Fourier series along the circle parameter.
The results are Fourier coefficients for every coordinate function,
which smoothly depend on the parameter of the interval. The entries
of the coordinate matrices corresponding to the coordinate functions
then can be defined by discretizing the Fourier coefficient, i.e.
a matrix entry is the value of a Fourier coefficient at a specific
coordinate of the interval, which coordinate depends on the indices
of the matrix entry.

Chapter 3 relates to direct sums of fuzzy spaces. We show that direct
sums of immersed cylinder fuzzy spaces can be related to matrix-valued
functions on a cylinder. We furthermore investigate unitary transformations
that naturally can be applied to matrix-valued functions. Such unitary
transformations also can be regularized into unitary transformations
of the fuzzy space. As the unitary transformations are already present
in the classical regime, their regularizations are examples of transformations
that do not correspond to smooth symplectomorphism.

In chapter 4 we generalize the construction of immersed cylinder fuzzy
spaces to general matrix-valued functions. We show that when commuting
matrix-valued functions are used, a classical limit can be defined.
Usually, the classical limit of a fuzzy space is defined via a series
of fuzzy spaces, which have the same properties, for example which
are the representations of the same algebra, and which in a sense
converge to a symplectic manifold, see \cite{Arnlind:2012} for a
rigorous definition. For fuzzy spaces defined by matrix algebras,
this is usually achieved by letting the size $N$ of the matrices
grow to infinity. 

For the type of fuzzy spaces developed in chapter 4, we show that
in the large $N$ limit, the regularization of the product of two
matrix-valued functions converges to the product of the quantization
of these functions, when the two functions commute up to first order
in $\frac{1}{N}$. Additionally, a generalization for a Poisson structure
for matrix-valued functions is presented in this context. As will
be shown, matrix-valued functions commuting up to first order can
be interpreted as coordinate patches of a manifold.

In chapter 5, a general construction is provided, which makes it possible
to interpolate two fuzzy spaces smoothly with each other. With this
interpolation method, a fuzzy string vertex with classical limit,
i.e. a surface that interconnects three circles with each other, is
constructed. Such a string vertex can be used to construct surfaces
of higher genus. For example, two string vertices can be concatenated
into a torus. In such a way, fuzzy surfaces having as classical limit
a surface of any genus can be constructed.

In chapter 6 we compare the new type of fuzzy spaces with already
known fuzzy spaces, like the fuzzy cylinder and the fuzzy torus. The
known fuzzy spaces are brought by a non-commutative coordinate transformation,
i.e. by a mapping of the coordinate matrices into other matrices,
into a form, which is suitable for a comparison. It is shown that
these coordinate transformed fuzzy spaces have the same features and
that also some kind of fuzzy string vertex is present.

\section{Immersed cylinder fuzzy spaces}

In the beginning, let us give a more rigorous definition of a fuzzy
space, which is used herein. A fuzzy space is a set of $d$ Hermitian
$N\times N$-matrices $\hat{X}^{\alpha},$ $a=1,\dots,d$, which can
be interpreted as the quantized embedding functions $x^{\alpha}$
of a classical manifold immersed in $\mathbb{R}^{d}$. 

Unitary transformations applied to the coordinate matrices and more
general to matrices of the algebra generated by these matrices 
\begin{equation}
\hat{X}^{\alpha}\longrightarrow\hat{U}\hat{X}^{\alpha}\hat{U}^{\dagger}\label{eq:unitary_trafo}
\end{equation}
correspond to symplectomorphic coordinate transformations. As a Hermitian
matrix always can be diagonalized with a unitary transformation, we
can fix this symmetry (at least partially up to permutations) be demanding
that one of the matrices, such as the last one $\hat{X}^{d}$ is diagonal.

In the following we often will restrict to a three dimensional fuzzy
space with three Hermitian $N\times N$-matrices $\hat{X}$, $\hat{Y}$
and $\hat{Z}$ and assume that $\hat{Z}$ is diagonal.

\subsection{Classical limit}

The fuzzy spaces presented below are more precisely series of matrix
algebras, numbered by the size $N$ of the matrices $\hat{X}_{N}^{\alpha}$.
The entries of the matrices $\hat{X}_{N}^{\alpha}$ are determined
from specific values of the Fourier components of a function $x^{\alpha}$
on a manifold immersed in $\mathbb{R}^{d}$. (see (\ref{eq:quant_fuzzy_cyl})
and (\ref{eq:matrix_reg}) below). These series, are matrix regularizations
according to \cite{Arnlind:2012}, i.e. the classical limit of these
series of matrix algebras is the space of functions on the manifold
immersed in $\mathbb{R}^{d}$.

The definition of a matrix regularization according to \cite{Arnlind:2012}
is the following: Let $Q_{N}$ be a series of linear maps from a space
of functions on a manifold $M$ with functions $f$ (here functions
on a submanifold of $\mathbb{R}^{d}$) to Hermitian $N\times N$-matrices
$Q_{N}(f)$. Then the $Q_{N}$ are a matrix regularization, when
\begin{enumerate}
\item The matrices $Q_{N}(f)$ converge in matrix norm, i.e. for a function
$f$
\begin{equation}
\lim_{N\rightarrow\infty}\left\Vert Q_{N}(f)\right\Vert <\infty\label{eq:conv_criteria_1}
\end{equation}
(for example, $\left\Vert \hat{f}\right\Vert =\max_{1<n<N}\sum_{n=1}^{N}\left|f_{nm}\right|$for
a matrix $\hat{f})$.
\item The matrix multiplication converges to the ordinary product of functions,
i.e. for For two functions $f$, $g$
\begin{equation}
\lim_{N\rightarrow\infty}\left\Vert Q_{N}(f)Q_{N}(g)-Q_{N}(fg)\right\Vert =0\label{eq:conv_criteria_2}
\end{equation}

\item For differentiable functions $f$, $g$ the commutator converges to
a Poisson bracket
\begin{equation}
\lim_{N\rightarrow\infty}\left\Vert -i\beta(N)[Q_{N}(f),Q_{N}(g)]-Q_{N}(\{f,g\})\right\Vert =0\label{eq:conv_criteria_3}
\end{equation}
i.e. the manifold is a Poisson manifold. The function $\beta(N)$
should have the property to converge to a constant for $N\rightarrow\infty$.
\end{enumerate}
Due to the last property, a matrix regularization can be seen as a
special kind of quantization in the sense of quantum mechanics, at
least, when the Poisson bracket is none-degenerate. In this case,
the manifold becomes a symplectic manifold and functions on the symplectic
manifold are mapped by the regularization to operators in a way that
the commutator of two functions is nearly the regularization of the
Poisson bracket induced by the symplectic structure.

We also introduce ``convergence within border'' and ``equivalence
within border'' for matrices. To define this, let $\delta$ be small
compared to $N$ and let $\hat{f}_{\delta}$ be the matrix, where
the outer border of size $\delta$ has been set to $0,$ i.e. $\hat{f}_{\delta,nm}=\hat{f}_{nm}$
for $\delta<n,m<N-\delta$ and $\hat{f}_{\delta,nm}=0$ for $n,m\leq\delta$
and $N-\delta\leq n,m$. 

We say that a matrix $\hat{f}_{N}$ converges within the border $\delta$,
when $\lim_{N\rightarrow\infty}\left\Vert \hat{f}_{N,\delta}\right\Vert $
converges. Equivalently, we also can use a weaker kind of matrix norm,
which only accounts for matrix entries within the border, such as
$\left\Vert \hat{f}\right\Vert =\max_{\delta<n<N-\delta}\sum_{n=\delta}^{N-\delta}\left|f_{nm}\right|.$
Analogously, we say that two matrices $\hat{f}$ and $\hat{g}$ are
equivalent within the border $\delta$, when $\hat{f}_{\delta}$=$\hat{g}_{\delta}$.
With such definitions, we are able to account for fuzzy spaces that
in a sense have a manifold with border as classical limit.

\subsection{Generalized fuzzy cylinder}

A simple example of a fuzzy space is the fuzzy cylinder (see \cite{Steinacker:2011,Sykora:2016}).
Generalizations of the fuzzy cylinder will play an important role
in the following. The fuzzy cylinder is based on a mapping from functions
to Toeplitz matrices, i.e. matrices in which each descending diagonal
from left to right is constant. 

A basis for Toeplitz matrices are $N\times N$-matrices $\hat{e}_{a,N}$
with only $1$ on the $a$.th diagonal, i.e. 
\begin{equation}
\hat{e}_{a,N}=\sum_{n}\ensuremath{\left|n\right\rangle \left\langle n+a\right|}
\end{equation}
where $a$ can be an integer between $-N+1$ and $N-1$. For example,
$\hat{e}_{0}$ is the unit matrix. On the other hand let $f(\varphi)=\sum f_{n}e^{in\varphi}$
be the Fourier series of the function $f$. The mapping from functions
to matrices via 
\begin{equation}
\hat{f}=Q_{N}(f)=\sum_{n=-N+1}^{N-1}f_{n}\hat{e}_{n,N}\label{eq:quant_fuzzy_cyl}
\end{equation}
 preserves multiplication of functions for $N\rightarrow\infty$,
since $Q_{\infty}(e^{in\varphi})Q_{\infty}(e^{im\varphi})=\hat{e}_{n}\hat{e}_{m}=\hat{e}_{n+m}=Q_{\infty}(e^{i(n+m)\varphi})$
and therefore
\begin{equation}
Q_{\infty}(f)(Q_{\infty}(g)=Q_{\infty}(fg)
\end{equation}
for two functions $f$ and $g$. The linear operators $Q_{N}$ are
a matrix regularization as defined above. In the remaining, we will
often indicate functions in the function space by its arguments, such
as $f(\varphi)$, and the corresponding regularized matrices with
a hat, such as $\hat{f}$.

The generalized fuzzy cylinder is defined in the following way. Let
$x(\varphi)$ and $y(\varphi)$ be two functions that define a closed
curve in $\mathbb{R}^{2}$ and let $x(\varphi)=\sum x_{n}e^{in\varphi}$
and $y(\varphi)=\sum y_{n}e^{in\varphi}$ be their Fourier series.
Then the matrices replacing the coordinate functions are defined via
\begin{eqnarray}
\hat{x}_{N} & = & Q(x)=\sum_{n=-N+1}^{N-1}x_{n}\hat{e}_{n,N}\nonumber \\
\hat{y}_{N} & = & Q(y)=\sum_{n=-N+1}^{N-1}y_{n}\hat{e}_{n,N}\label{eq:fuzzy_cylinder}\\
\hat{z}_{N} & = & \sum_{n=1}^{N}\beta\frac{n}{N}\left|n\right\rangle \left\langle n\right|\nonumber 
\end{eqnarray}
Since $\hat{x}_{N}$ and $\hat{y}_{N}$ are Toeplitz-matrices, they
commute, i.e. $[\hat{x}_{N},\hat{y}_{N}]=0.$ When we further assume
that $x_{n}=0$ and $y_{n}=0$ for $n>\delta$, where $\delta$ is
small compared to $N$, then the commutators with $\hat{z}$ fulfill
\begin{equation}
[\hat{z}_{N},\hat{x}_{N}]=Q\left(-i\beta\partial_{\varphi}x(\varphi)\right),\ [\hat{z_{N}},\hat{y}_{N}]=Q\left(-i\beta\partial_{\varphi}y(\varphi)\right)
\end{equation}
within the border $\delta$. For $N\rightarrow\infty$ the algebra
generated by the matrices converge in the sense of (\ref{eq:conv_criteria_1}),
(\ref{eq:conv_criteria_2}) and (\ref{eq:conv_criteria_3}) within
the border $\delta$ to an algebra of functions on a submanifold of
$\mathbb{R}^{3}$, which is the product of the curve defined by $x(\varphi)$
and $y(\varphi)$ times the interval $[0,\beta]$. This manifold has
the natural Poisson-bracket 
\begin{equation}
\{f,g\}=\partial_{\varphi}f\partial_{z}g-\partial_{z}f\partial_{\varphi}g
\end{equation}
It is also possible to start in (\ref{eq:fuzzy_cylinder}) with $\hat{x}_{\infty}$
and $\hat{y}_{\infty}$ and use $\hat{z}_{\beta}=\sum_{n}\beta n\left|n\right\rangle \left\langle n\right|$.
When we let $\beta$ got to $0$, the classical limit is the curve
defined by $x(\varphi)$ and $y(\varphi)$ times $\mathbb{R}$. However,
in this case, the definitions (\ref{eq:conv_criteria_1}), (\ref{eq:conv_criteria_2})
and (\ref{eq:conv_criteria_3}) have to be extended to operators on
a Hilbert space.

\subsection{Immersed cylinder}

For a further generalization of the fuzzy cylinder, we start again
with a function defined by a Fourier series, which however depends
on a second parameter 
\begin{equation}
f(q,\varphi)=\sum_{n\in\mathbb{N}}f_{n}(q)e^{in\varphi}\label{eq:fourier_single_valued}
\end{equation}
with $q\in[q_{1},q_{2}]$ a finite interval and $\varphi\in[0,2\pi]$.
We note that, when the Fourier series is truncated at a specific $n$,
for example for $|n|<\delta$ the function $f$ is approximated by
the truncated Fourier series. In the following it is assumed that
$f_{n}=0$ for $\left|n\right|\geq\delta$ for a constant $\delta$
that is small compared to $N$. 

A matrix regularization $Q_{N}(f)$ for the function $f$ is defined
with the $N\times N$-matrix 
\begin{equation}
Q_{N}(f)=\hat{f}=\sum_{n,m=1}^{N}f_{m-n}\left(q(n,m)\right)\hat{e}_{nm}\label{eq:def_regularization}
\end{equation}
where $\hat{e}_{nm}=\left|n\right\rangle \left\langle m\right|$.
This definition can be extended to ``infinite dimensional'' matrices,
i.e. $N=\infty$. Also with $N=\infty$, criteria (\ref{eq:conv_criteria_1})
is fulfilled, due to the $\delta$-condition. When $f$ does not depend
on $q$, definition (\ref{eq:def_regularization}) reduces to the
regularization (\ref{eq:quant_fuzzy_cyl}) of the generalized fuzzy
cylinder.

$q(n,m)$ is a discretizing function, which assigns the matrix entry
$\hat{e}_{nm}$ to a point in the interval of $q$. $q(n,m)$ should
have to following properties:
\begin{itemize}
\item It should fill out the interval for $q$: $q(0,0)=q_{1}$, $q(N,N)=q_{2}$
(as for large $N$ $q(0,0)\approx q(1,1)$, we can take $q(0,0)$
instead of $q(1,1)$, which renders the formulas for the examples
for $q(n,m)$ below less complex).
\item It should depend ``smooth'' on its arguments, i.e $q(m,p)=q(m,n)+\frac{\beta}{N}(p-n)$
for $p-n$ small compared to $N$ and some constant $\beta$ that
can depend on $m$,$n$ and $p$. Analogously, $q(p,n)=q(m,n)+\frac{\beta}{N}(p-m)$
for $p-m$ small compared to $N$ . 
\item Examples are $q(n,m)=\frac{q_{2}-q_{1}}{2N}(m+n)+q_{1}$ (in this
case $\beta=\frac{1}{2}$) or $q(n,m)=\frac{q_{2}-q_{1}}{N}n+q_{1}$
(in this case $\beta=1$ for the left argument and $\beta=0$ for
the right argument).
\end{itemize}
When the function $f$ is real-valued, the resulting matrix $Q_{N}(f)$
is Hermitian for a symmetric discretizing function $q(m,n)=q(n,m)$,
since for a real-valued function, the coefficients of the Fourier
series have the property that $\bar{f}_{n}=f_{-n}$.

Analogously to \cite{Shimada:2003}, we can now evaluate the matrix
product of two matrices $\hat{f}$ and $\hat{g}$ as defined in (\ref{eq:def_regularization}),
which have one times differentiable Fourier coefficient functions
$f_{n}(q)$ and $g_{n}(q)$
\begin{eqnarray}
(\hat{f}\hat{g})_{nm} & = & \sum_{1<p<N}f_{p-n}\left(q(n,p)\right)g_{m-p}\left(q(p,m)\right)\nonumber \\
 & = & \sum_{\left|n-p\right|<\delta,\left|p-m\right|<\delta}f_{p-n}\left(q(n,m)+\frac{\beta}{N}(p-m)\right)g_{m-p}\left(q(n,m)+\frac{\beta}{N}(p-n)\right)\nonumber \\
 & = & \sum_{\left|n-p\right|<\delta,\left|p-m\right|<\delta}f_{p-n}g_{m-p}\:\:\:\:\:\:\:\:\:\:\nonumber \\
 &  & \:\:\:\:\:\:\:\:\:\:+\frac{\beta}{N}\left(f_{p-n}(p-n)\frac{\partial g_{p-m}}{\partial q}+(p-m)\frac{\partial f_{n-p}}{\partial q}g_{m-p}\right)+\mathcal{O}\left(\frac{1}{N^{2}}\right)\label{eq:ev_immersed_cylinder}
\end{eqnarray}
where in the last step all functions are evaluated at $q=q(n,m)$.
Since 
\begin{equation}
-i\frac{\partial f}{\partial\varphi}=\sum_{n}nf_{n}(q)e^{in\varphi}\label{eq:formal_derivative}
\end{equation}
the last equation is equivalent to
\begin{eqnarray}
Q_{N}(f)Q_{N}(g) & = & Q_{N}(fg)+i\frac{\beta}{N}Q\left(\frac{\partial f}{\partial\varphi}\frac{\partial g}{\partial q}-\frac{\partial g}{\partial\varphi}\frac{\partial f}{\partial q}\right)+\mathcal{O}\left(\frac{1}{N^{2}}\right)\nonumber \\
 & = & Q_{N}(fg)+i\frac{\beta}{N}Q_{N}\left(\{f,g\}\right)+\mathcal{O}\left(\frac{1}{N^{2}}\right)\label{eq:approx_product}
\end{eqnarray}
It follows that the commutator of such regularized functions can be
approximated by the Poisson bracket of the functions
\begin{equation}
[Q_{N}(f),Q_{N}(g)]=i\frac{2\beta}{N}Q_{N}\left(\{f,g\}\right)+\mathcal{O}\left(\frac{1}{N^{2}}\right)
\end{equation}

Thus, all three criteria (\ref{eq:conv_criteria_1}), (\ref{eq:conv_criteria_2}),
(\ref{eq:conv_criteria_3}) above are fulfilled. Note that the Poisson
bracket of the classical limit is $\{q,\varphi\}=-1$, which also
follows from $[\hat{q},Q_{N}(e^{i\varphi})]=\frac{1}{N}Q_{N}(e^{i\varphi})$.

When the coefficient functions $f_{n}(q)$ and $g_{n}(q)$ are only
continuous, one can show that $Q(f)Q(g)=Q(fg)+\mathcal{O}\left(\frac{1}{N}\right)$
and $[Q(f),Q(g)]=\mathcal{O}\left(\frac{1}{N}\right)$, i.e. the criteria
(\ref{eq:conv_criteria_1}), (\ref{eq:conv_criteria_2}) are still
fulfilled.

From (\ref{eq:formal_derivative}) one sees that the Fourier coefficients
$f_{n}$ for $n\rightarrow\infty$ need to go to $0$ faster than
$\frac{1}{n}$, which is the case for functions with $f_{n}=0$ for
$\left|n\right|\geq\delta$. However, also other functions, like functions
twice differentiable in $\varphi$ and specific piece wise continuous
functions (but not all) have this property. In (\ref{eq:ev_immersed_cylinder})
only the differentiability in $q$ was used. As long as the Fourier
coefficients go to $0$ as fast such that (\ref{eq:ev_immersed_cylinder})
is fulfilled, (\ref{eq:def_regularization}) defines a matrix regularization.
Thus, it is possible to start with functions $x(q,\varphi)$ and $y(q,\varphi)$,
which do not define a closed curve for a specific $q$, but which
only have Fourier coefficients, which go to $0$ fast enough. In this
case, it would be better to talk of a fuzzy immersed strip. The same
applies to the generalized cylinder as described in the previous section.

Up to now, we have ignored, what happens at the borders $q=q_{1}$
and $q=q_{2}$. Due to the discretizing function $q(n,m)$, which
maps values at the border of the $q$-interval to the border of the
matrices, we have to look at $n<\delta$ and $N-n<\delta$. We see
that formula (\ref{eq:approx_product}) is only valid within the border
$\delta.$ This problem can be avoided, when the coefficient functions
$f_{n}(q)$ and $g_{n}(q)$ go to zero at the border, i.e. when $f_{n}(q_{1})=f_{n}(q_{2})=0$.
This can be seen by embedding a matrix of size $N$ in a bigger zero
matrix, resulting in a border of zeros surrounding the original matrix.
The continuation of functions $f_{n}(q)$, which do not go to zero
at the border, into this border would result in non-continuous extended
functions, which are not differentiable. 

When we demand that $f(q_{1},\varphi)=f(q_{2},\varphi)=0$, the submanifolds
that can be regularized with $Q_{N}$ have to be immersions of a cylinder
into $\mathbb{R}^{d}$ with ends shrunk to a point. When the submanifold
has a border, the $Q_{N}$ only converge within border.

When the target space is $\mathbb{R}^{3}$, a two-dimensional immersed
submanifold can be para\-metrized by the three functions
\begin{eqnarray}
x(q,\varphi) & = & \sum_{n\in\mathbb{N}}x_{n}(q)e^{in\varphi}\nonumber \\
y(q,\varphi) & = & \sum_{n\in\mathbb{N}}y_{n}(q)e^{in\varphi}\label{eq:immersed_fuzzy_cylinder}\\
z(q,\varphi) & = & z(q)\nonumber 
\end{eqnarray}

The fuzzy space defined by matrix regularization of these functions
with $N\times N$-matrices is then 
\begin{eqnarray}
\hat{x} & = & \sum_{n,m=1}^{N}x_{m-n}\left(q(n,m)\right)\hat{e}_{nm}\nonumber \\
\hat{y} & = & \sum_{n.m=1}^{N}y_{m-n}\left(q(n,m)\right)\hat{e}_{nm}\\
\hat{z} & = & \sum_{n=1}^{N}z(q(n,n))\hat{e}_{nn}\nonumber 
\end{eqnarray}

\subsection{Example: from circle to 8\label{sub:Example:-from-circle}}

In this section we give an example based on the fuzzy immersed cylinder
that later on will be generalized to a fuzzy manifold having the structure
of a string vertex. Here, we construct the part of the vertex, which
at one end with respect to the $z$-direction has a circular cross-section
and transforms along the $z$-direction into a cross-section similar
to an 8. The center crossing of the eight is the point, where the
string vertex splits into two cylinders.

To define the fuzzy \textquotedbl{}circle-to-eight''-space let us
start with two coordinate functions in polar coordinates, i.e. $x(z,\varphi)=r(z,\varphi)\cos\varphi$
and $y(z,\varphi)=r(z,\varphi)\sin\varphi$ with $r(z,\varphi)=r_{1}(z)+r_{2}(z)\cos2\varphi$.
This is equivalent to

\begin{eqnarray}
x(z,\varphi) & = & \left(r_{1}(z)+\frac{r_{2}(z)}{2}\right)\cos\varphi+\frac{r_{2}(z)}{2}\cos3\varphi\nonumber \\
y(z,\varphi) & = & \left(r_{1}(z)-\frac{r_{2}(z)}{2}\right)\sin\varphi+\frac{r_{2}(z)}{2}\sin3\varphi\label{eq:circle-to-eight}
\end{eqnarray}
With $r_{1}(z)=1$ and $r_{2}$ continuously increasing between $z_{1}$
and $z_{2}$ and $r_{2}(z_{1})=0$ and $r_{2}(z_{2})=1$ the corresponding
immersed manifold has a transition from a circle to a 8 as shown to
the left of Fig. 1.

\begin{figure}
\includegraphics[scale=0.5]{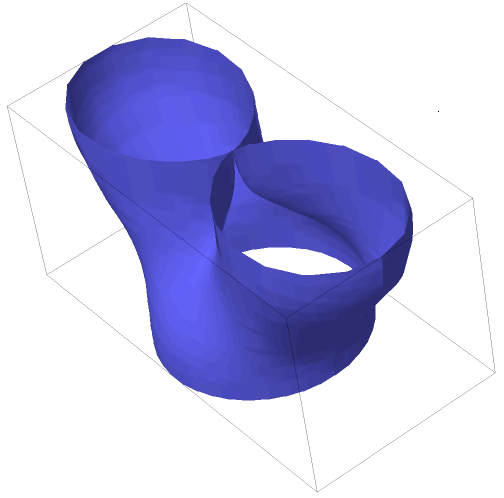}\includegraphics[scale=0.4]{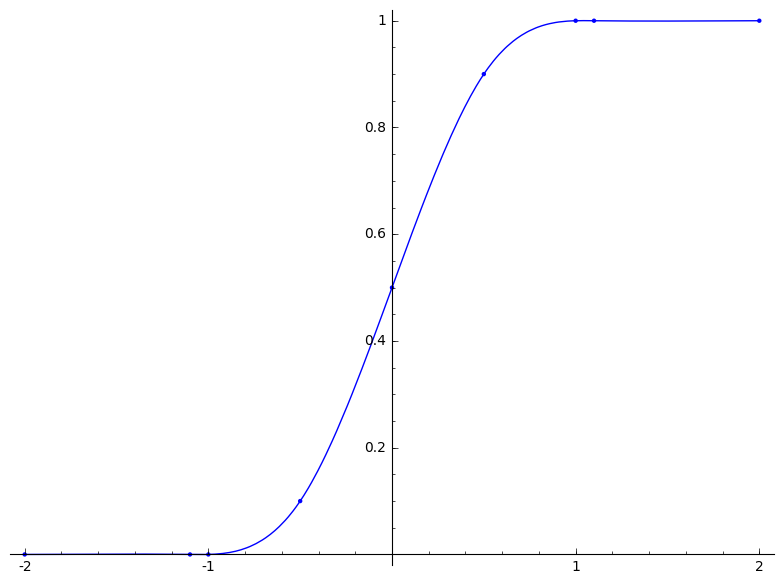}\caption{From circle to eight}
\end{figure}

To fix the functions $r_{1}$ and $r_{2}$ between $z_{1}$ and $z_{2}$,
we need a continuously increasing, differentiable function $h(q)$
with $h(q<-1)=0$ and $h(q>1)=1$. For the figures and matrices in
this article, this function is always represented with a cubic spline
interpolating the points $(-1,0)$, $(-0.5,0.1)$, $(0,0.5)$$,(0.5,0.9)$
and $(1,1)$ with further support points at $h=0$ and $h=1$ outside
of $-1$ and $1$. The spline $h$ with it support points (and the
function $r_{2}$, which is here equivalent to $h$) is shown to the
right of Fig. 1.

A matrix regularization according to (\ref{eq:def_regularization})
for the manifold shown in Fig. 1 is
\begin{equation}
\hat{z}=\sum_{n}\frac{n}{N}\hat{e}_{nn}
\end{equation}
\begin{equation}
\hat{x}=\sum_{n}\left(1+\frac{1}{2}h(-1+\frac{2n}{N})\right)\frac{1}{2}\left(\hat{e}_{n,n+1}+\hat{e}_{n+1,n}\right)+\frac{1}{2}h(-1+\frac{2n}{N})\frac{1}{2}\left(\hat{e}_{n,n+3}+\hat{e}_{n+3,n}\right)
\end{equation}
\begin{equation}
\hat{y}=\sum_{n}\left(1-\frac{1}{2}h(-1+\frac{2n}{N})\right)\frac{i}{2}\left(\hat{e}_{n,n+1}-\hat{e}_{n+1,n}\right)+\frac{1}{2}h(-1+\frac{2n}{N})\frac{i}{2}\left(\hat{e}_{n,n+3}-\hat{e}_{n+3,n}\right)
\end{equation}
Here, $z=1+2q$ and $q$ is discretized with $-1+\frac{2n}{N}$. Again,
as the ends of the circle-to-eight manifold are not closed, the algebra
defined by these matrices only converges within border in the sense
of (\ref{eq:conv_criteria_1}) to (\ref{eq:conv_criteria_3}). Completely
converging matrices can be formed by providing the ends with ``caps'',
for example by fading out the Fourier coefficients in an interval
larger than $[-1,1]$ and extending the matrices correspondingly.

Fig. 2 shows a visualization of the matrices $\hat{x}$, $\hat{y}$
and $\hat{z}$ with $N=30$ in the form of ``dot matrix diagrams''.
Every matrix is represented by a matrix of circles. Every circle represents
a matrix entry and has a radius proportional to the root of the absolute
value of the respective matrix entry. Matrix entries equal to $0$
are depicted with a small point, such that the rows and columns can
be better identified. We will use these dot matrix diagrams in the
following for more complicated examples. They represent an easy way
to visually compare properties of matrices with each other.

\begin{figure}
\includegraphics[scale=0.3]{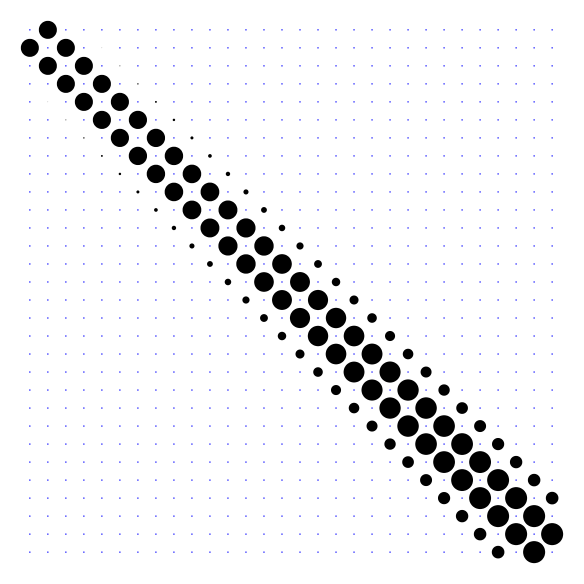} \includegraphics[scale=0.3]{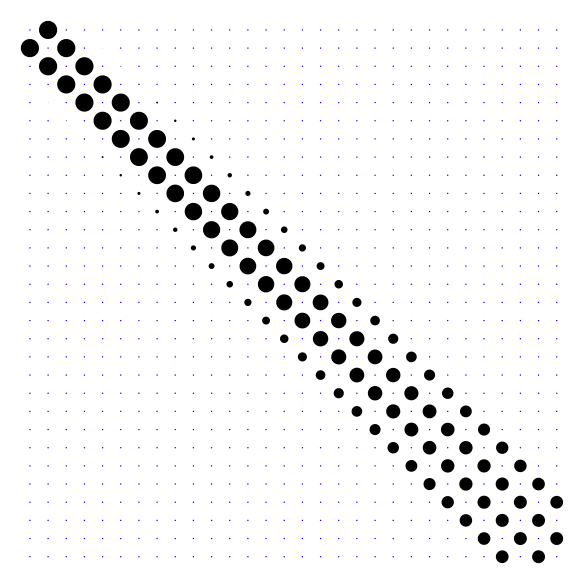}
\includegraphics[scale=0.3]{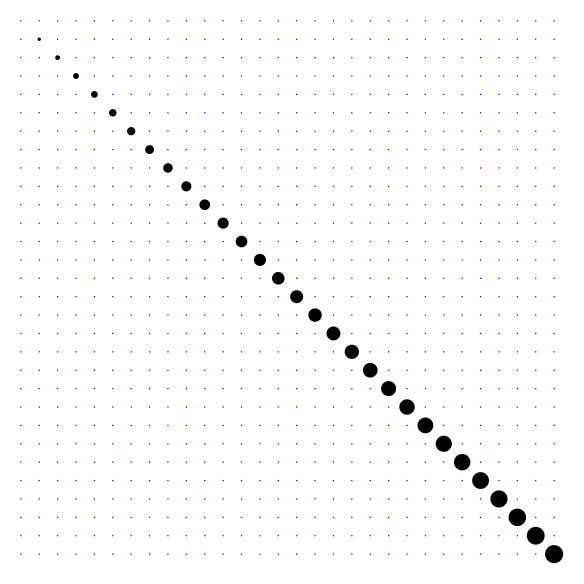}\caption{Visualization for circle-to-eight matrices $\hat{x}$, $\hat{y}$
and $\hat{z}$}
\end{figure}

\section{Direct sums}

For two manifolds (that can be immersed in the same space), the space
of functions on the two manifolds is the direct sum of the function
spaces of the manifolds. Thus, if one wants to compose a fuzzy space
of two other fuzzy spaces, the corresponding operation is the direct
sum of matrices. When $X_{1}^{\alpha}$ and $X_{2}^{\alpha}$ are
two sets of matrices that are matrix regularizations of two submanifolds
in the same space, the matrices $X^{\alpha}=X_{1}^{\alpha}\oplus X_{2}^{\alpha}$
represent a matrix regularization of both submanifolds:
\begin{equation}
X^{\alpha}=\left(\begin{array}{cc}
X_{1}^{\alpha}\\
 & X_{2}^{\alpha}
\end{array}\right)\label{eq:direct_sum}
\end{equation}
In the following, we will see that a matrix regularization of a direct
sum of fuzzy spaces can be seen as a matrix regularization of diagonal
matrix-valued functions. By generalizing to arbitrary matrix-valued
functions, we are able to describe a new class of fuzzy spaces in
the next chapter.

\subsection{Matrix regularizations of direct sums}

In general, let us assume that we have two (or more general $S$)
matrices $\hat{f}_{a}$, which are regularizations of the functions
$f_{a}(q,\varphi).$ The direct product of these matrices is the matrix
$\hat{F}$ with $F_{ab,nm}=f_{a,nm}\delta_{ab}$, where we have introduced
a double-index notation with $a,b=1\ldots S$ and $i,j=1\ldots N$.
For example, with $S=2$ 
\begin{equation}
\hat{F}=\left(\begin{array}{cccccc}
f_{1,11} & f_{1,12} & \cdots\\
f_{1,21} & f_{1,22} & \cdots\\
\vdots & \vdots & \ddots\\
 &  &  & f_{2,11} & f_{2,12} & \cdots\\
 &  &  & f_{2,21} & f_{2,22} & \cdots\\
 &  &  & \vdots & \vdots & \ddots
\end{array}\right)=\sum_{n,m}\left(\begin{array}{cc}
f_{1,nm}\hat{e}_{nm}\\
 & f_{2,nm}\hat{e}_{nm}
\end{array}\right)\label{eq:pure_direct_sum}
\end{equation}
Due to the invariance of a fuzzy space with respect to permutations
(being a special kind of unitary transformation), we also can exchange
the role of the ``inner'' and the ``outer'' indices
\begin{equation}
\hat{F}'=\hat{P}\hat{F}\hat{P}=\left(\begin{array}{ccccc}
f_{1,11} &  & f_{1,12} &  & \cdots\\
 & f_{2,11} &  & f_{2,12} & \cdots\\
f_{1,21} &  & f_{1,22} &  & \cdots\\
 & f_{2,21} &  & f_{2,22} & \cdots\\
\vdots & \vdots & \vdots & \vdots & \ddots
\end{array}\right)=\sum_{i,j}\left(\begin{array}{cc}
f_{1,nm}\\
 & f_{1,nm}
\end{array}\right)\hat{e}_{nm}\label{eq:z-ordered}
\end{equation}
where $\hat{P}$ is a suitable permutation. The matrix $\hat{F}'$
can be seen as a regularization of a matrix-valued function $F$
\begin{equation}
\hat{F}'=Q\left(F\right)\ \mathrm{with}\ F(q,\varphi)=\left(\begin{array}{cc}
f_{1}(q,\varphi)\\
 & f_{2}(q,\varphi)
\end{array}\right)\label{eq:diagonal_matrix-valued function}
\end{equation}
From now on, we will consider $\hat{F}'$ and $\hat{F}$ as equivalent.
In summary, we have transformed the regularization of the two real-valued
functions $f_{1}$ and $f_{2}$, which are defined on the cylinder
(or more general strip) $[q_{l},q_{r}]\times[0,2\pi]$ to a regularization
of one matrix-valued function $F$ defined on this cylinder.

Due to (\ref{eq:z-ordered}), the borders of the matrices $\hat{f}_{a}$
also have been shifted to the border of the matrix $\hat{F}$. Thus,
when we have matrices converging in border then also their direct
product permuted according to (\ref{eq:z-ordered}) converges in border.

\subsection{Constant unitary transformations}

The coordinate functions $x_{1}^{\alpha}(q,\varphi)$ and $x_{2}^{\alpha}(q,\varphi)$
used for defining the matrices of two fuzzy spaces by regularization
can be composed into matrices $X^{\alpha}(q,\varphi)$ according to
(\ref{eq:diagonal_matrix-valued function}). When a constant unitary
transformation $U$, i.e. a unitary transformation independent of
$q$ and $\varphi$ is applied to the matrix-valued functions $X^{\alpha}(q,\varphi)$,
this is equivalent to applying the unitary transformation 
\begin{equation}
\hat{U}=\sum_{n}U\hat{e}_{nn}\label{eq:constant_unitary}
\end{equation}
(which can be seen as the regularization of the matrix $U$) to the
corresponding regularization $\hat{X^{\alpha}}$(\ref{eq:direct_sum}).
In this case, multiplication of matrices commutes with regularization
\begin{equation}
Q(U^{\dagger}FU)=\hat{U}^{\dagger}\hat{F}\hat{U}
\end{equation}
On the other hand, when we start with a set of general matrix-valued
functions $X^{\alpha}(q,\varphi)\in M_{N}(\mathbb{C})$, such as 
\begin{equation}
X^{\alpha}(q,\varphi)=\left(\begin{array}{cc}
X_{11}^{\alpha}(q,\varphi) & X_{12}^{\alpha}(q,\varphi)\\
X_{21}^{\alpha}(q,\varphi) & X_{12}^{\alpha}(q,\varphi)
\end{array}\right)
\end{equation}
and the matrices $X^{\alpha}(q,\varphi)$ are simultaneously diagonalizable
with a constant unitary transformation, we can associate a direct
sum of fuzzy spaces to the matrix regularization. Due to to the matrix
form of the defining functions $X^{\alpha}(q,\varphi)$, a new class
of symmetry is present, namely constant $U(n)$-transformations, which
can be seen as a generalization of the trivial $U(1)$ symmetry of
single-valued functions. After defining a regularization for general
matrix-valued functions, we will also see that this symmetry can be
generalized to coordinate dependent $U(n)$-transformations.

An interesting case is, when one of the matrix-valued functions $Z(q,\varphi)$
is proportional to the unit matrix, i.e. $Z_{ab}=z(q,\varphi)\delta_{ab}$.
Fuzzy spaces based on direct sums have this property, when the same
diagonal matrix for the $z$-coordinate is used for both spaces. In
this case, a residual symmetry remains that can be used to transform
the other matrices $\hat{X}^{\alpha}.$ Every unitary transformation
$\hat{U}$ (\ref{eq:constant_unitary}) based on a constant unitary
transformations $U$ leaves the matrix $\hat{Z}$, see (\ref{eq:z-ordered}),
invariant. 

For example, we can consider the unitary transformation 

\begin{equation}
U=\frac{1}{\sqrt{2}}\left(\begin{array}{cc}
1 & -1\\
1 & 1
\end{array}\right)\label{eq:interlacing_unitary}
\end{equation}
which has the property that it mixes the diagonal and off-diagonal
elements of the matrix-valued function $F=\left(\begin{array}{cc}
f_{11} & f_{12}\\
f_{21} & f_{22}
\end{array}\right)$ 
\begin{equation}
U^{\dagger}FU=\frac{1}{2}\left(\begin{array}{cc}
(f_{11}+f_{22})+(f_{12}+f_{21}) & (f_{22}-f_{11})+(f_{12}-f_{21})\\
(f_{22}-f_{11})-(f_{12}-f_{21}) & (f_{11}+f_{22})-(f_{12}+f_{21})
\end{array}\right)
\end{equation}
Such a transformation will below become important for two symmetric
fuzzy spaces, which have the property that $x_{1}^{\alpha}(q,\varphi)=-x_{2}^{\alpha}(q,\varphi)$
for the coordinate functions except $z(q,\varphi)$. In this case,
the transformation results in a swap between diagonal entries and
off-diagonal entries.

\begin{equation}
F_{I}=U^{\dagger}\left(\begin{array}{cc}
-f\\
 & f
\end{array}\right)U=\left(\begin{array}{cc}
 & f\\
f
\end{array}\right)\label{eq:def_mirror_cyl_interlaced}
\end{equation}
We will call this ``interlacing'' of fuzzy spaces.

\subsection{Example: Two cylinders}

We will apply the preceding discussion to the direct sum of two generalized
fuzzy cylinders. In particular, the part of the string vertex, which
follows the circle-to-eight manifold (\ref{eq:circle-to-eight}),
consists of two cylinders, which move away from each other with increasing
$z$. In chapter 5, the fuzzy circle-to-eight space will be connected
with the two interlaced fuzzy cylinders, which are defined here.

Firstly, more generally, we consider two cylinders, which are aligned
along a center $\left((x_{i,0}(z),y_{i,0}(z)\right)$ ($i=1,2$ enumerates
the two cylinders) and which have two $z$-depend radii $r_{i,x}(z),\text{ }r_{i,y}(z)$
along the $x$-direction and the $y$-direction. A matrix regularization
for each cylinder is
\begin{eqnarray}
\hat{x_{i}} & = & \sum_{n}x_{i,0}\left(q_{n,n}\right)\hat{e}_{nn}+\frac{1}{2}r_{i,x}\left(q_{n,n+1}\right)\hat{e}_{n,n+1}+\frac{1}{2}r_{i,x}\left(q_{n+1,n}\right)\hat{e}_{n+1,n}\nonumber \\
\hat{y_{i}} & = & \sum_{n}y_{i,0}\left(q_{n,n}\right)\hat{e}_{nn}+\frac{1}{2}r_{i,y}\left(q_{n,n+1}\right)\hat{e}_{n,n+1}+\frac{1}{2}r_{i,x}\left(q_{n+1,n}\right)\hat{e}_{n+1,n}\label{eq:def_cylinder}\\
\hat{z} & = & \sum_{n}q_{n,n}\hat{e}_{nn}\nonumber 
\end{eqnarray}
The direct sum (\ref{eq:direct_sum}) of these two fuzzy spaces is
\begin{equation}
\hat{X}=\left(\begin{array}{cc}
\hat{x}^{1}\\
 & \hat{x}^{2}
\end{array}\right)\ \hat{Y}=\left(\begin{array}{cc}
\hat{y}^{1}\\
 & \hat{y}^{2}
\end{array}\right)\ \hat{Z}=\left(\begin{array}{cc}
q\\
 & q
\end{array}\right)
\end{equation}
The $2\times2$-matrix-valued functions (\ref{eq:diagonal_matrix-valued function})
for regularization are 
\begin{eqnarray}
X(q,\varphi) & = & \left(\begin{array}{cc}
x_{1,0}(q)+r_{1,x}(q)\cos\varphi\\
 & x_{2,0}(z)+r_{2,x}(z)\cos\varphi
\end{array}\right)\nonumber \\
Y(q,\varphi) & = & \left(\begin{array}{cc}
y_{1,0}(q)+r_{1,y}(q)\sin\varphi\\
 & y_{2,0}(q)+r_{2,y}(q)\sin\varphi
\end{array}\right)\label{eq:two_cylinders}\\
Z(q,\varphi) & = & \left(\begin{array}{cc}
q\\
 & q
\end{array}\right)\nonumber 
\end{eqnarray}

When the two cylinders are symmetric with respect to the $z$-axis,
this implies $x_{0}(q)=-x_{1,0}(q)=x_{2,0}(q)$ and $r_{x/y}(q)=-r_{1,x/y}(q)=r_{2,x/y}(q)$.
The interlaced form (\ref{eq:def_mirror_cyl_interlaced}) of the matrices
becomes 
\begin{eqnarray}
X_{I}(q,\varphi) & = & \left(\begin{array}{cc}
 & x_{o}(q)+r_{x}(q)\cos\varphi\\
x_{o}(q)+r_{x}(q)\cos\varphi
\end{array}\right)\nonumber \\
Y_{I}(q,\varphi) & = & \left(\begin{array}{cc}
 & r_{y}(q)\sin\varphi\\
r_{y}\sin\varphi
\end{array}\right)\label{eq:two_cyl_interlaced}\\
Z_{I}(q,\varphi) & = & \left(\begin{array}{cc}
q\\
 & q
\end{array}\right)\nonumber 
\end{eqnarray}
Fig. 3 shows dot matrix diagrams of matrix regularizations with $N=15$
for the two cylinders. The first row of diagrams shows the matrices
$\hat{X},$$\hat{Y}$and $\hat{Z}$ for the pure direct sum of the
two fuzzy cylinders, according to (\ref{eq:pure_direct_sum}). The
second row shows the matrices $\hat{X}',$$\hat{Y}'$and $\hat{Z}'$
after a permutation according to (\ref{eq:z-ordered}) The third row
shows the interlaced matrices according to (\ref{eq:two_cyl_interlaced})
adjoined with (\ref{eq:constant_unitary}) based on (\ref{eq:interlacing_unitary}).
If one compares Fig. 2 above for the circle-to-eight fuzzy space with
the third row of Fig. 3, one sees, that the same diagonals of the
matrices only have entries different from $0$. In chapter 5 we will
develop a method, how these two matrices for one coordinate can be
smoothly interconnected with each other.

\begin{figure}
\includegraphics[scale=0.3]{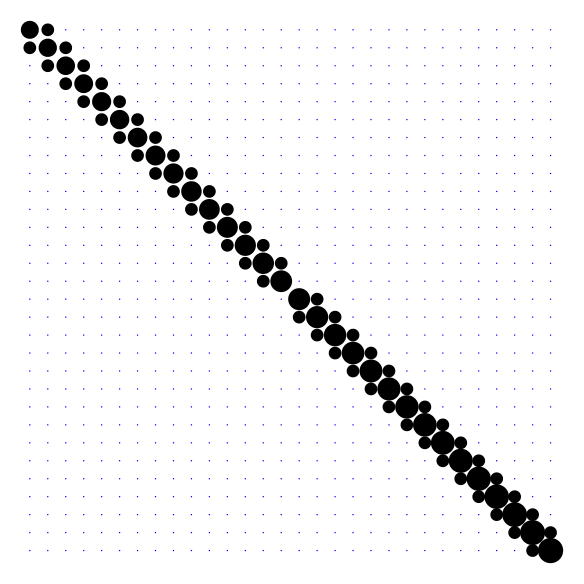} \includegraphics[scale=0.3]{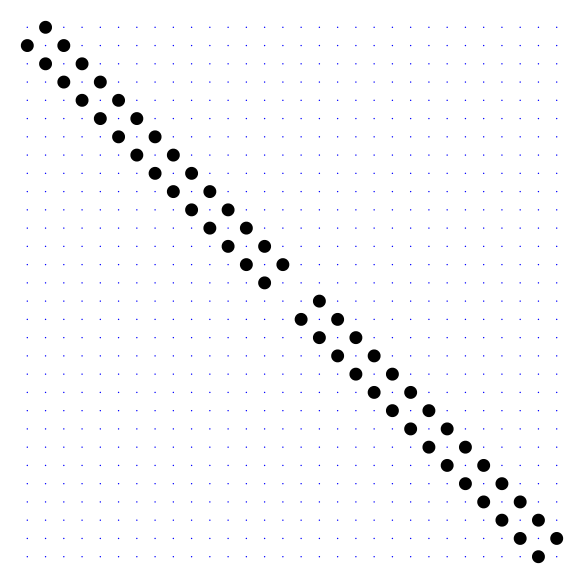}
\includegraphics[scale=0.3]{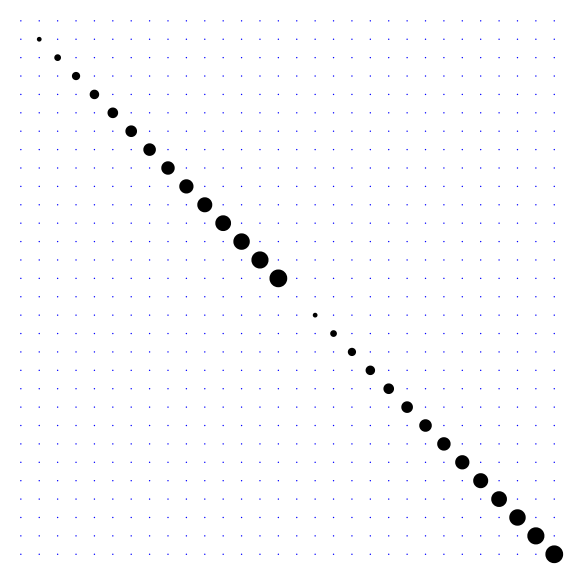}

\includegraphics[scale=0.3]{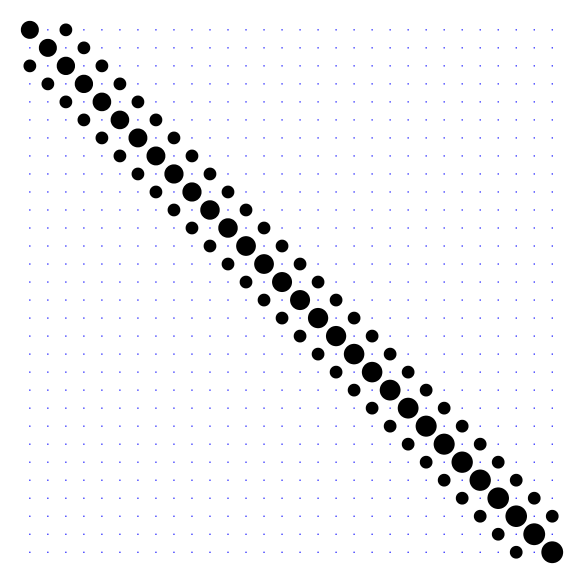} \includegraphics[scale=0.3]{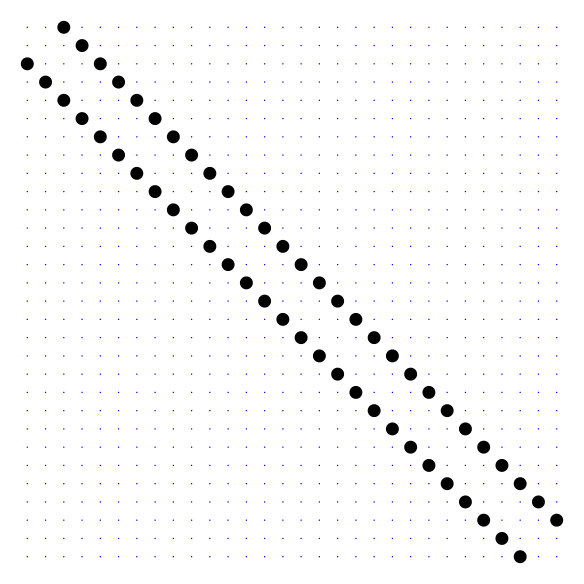}
\includegraphics[scale=0.3]{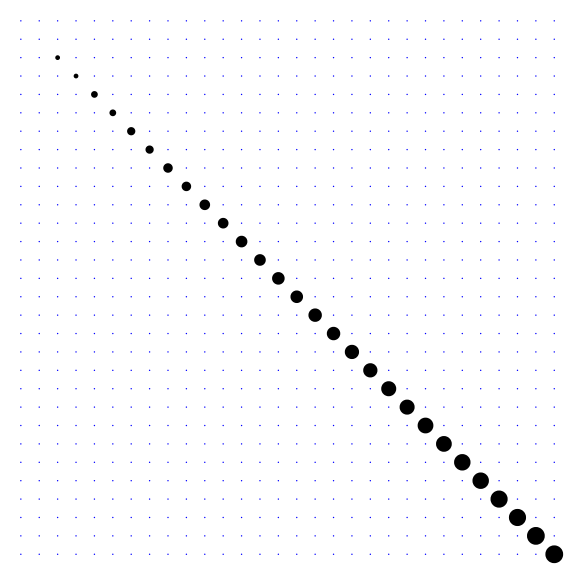}

\includegraphics[scale=0.3]{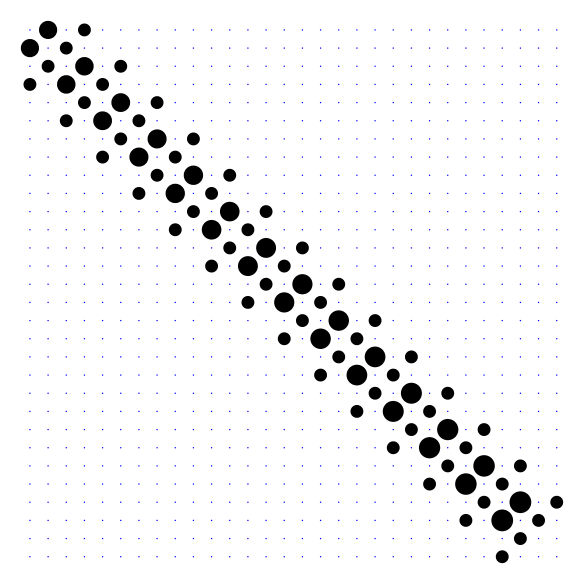} \includegraphics[scale=0.3]{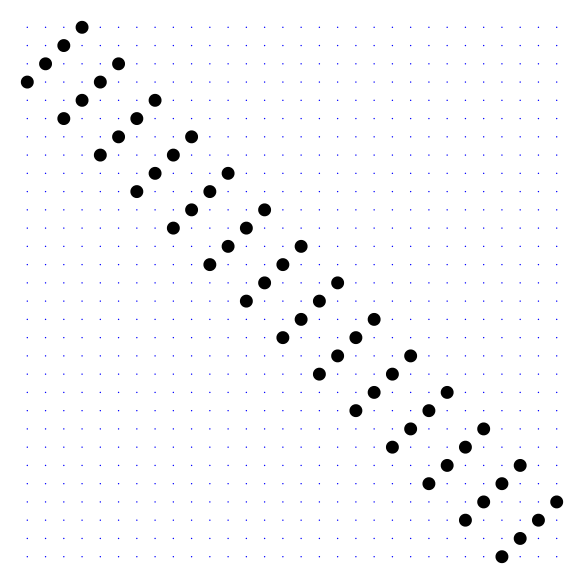}
\includegraphics[scale=0.3]{2cylZ2}

\caption{Visualization for two cylinder matrices: direct sum, z-ordered and
interlaced}
\end{figure}

\section{Fuzzy spaces based on matrix-valued functions}

In the previous chapter we have seen that the direct sum of two or
more fuzzy spaces, each of which is defined by a regularization of
coordinate functions, can be described with a regularization of diagonal
matrix-valued functions (\ref{eq:diagonal_matrix-valued function}).
In the following, we will generalize to fuzzy spaces defined by a
regularization of arbitrary matrix-valued functions. 

To give a precise definition of such a regularization, let $F_{ab}(q,\varphi)$
be a matrix-valued function with $a,b=1\ldots S$, which can be decomposed
into a Fourier series 
\begin{equation}
F_{ab}(q,\varphi)=\sum_{n}F_{ab,n}(q)e^{in\varphi}\label{eq:fourier_matrix_valued}
\end{equation}
Again $q\in[q_{1},q_{2}]$ is in a finite interval and $\varphi\in[0,2\pi]$.
A $N$-dimensional matrix regularization $Q_{N}(F)=\hat{F}$of $F$
is then 
\begin{equation}
\hat{F}_{ab,nm}=F_{ab,m-n}\left(q(n,m)\right)\label{eq:matrix_reg}
\end{equation}
with a discretizing function $q$ as defined above. Note that the
matrix $\hat{F}$ is actually a square matrix of dimensions $N$ times
$S$. When the discretizing function $q$ is symmetric and the function
$F(q,\varphi)$ is Hermitian, then the matrix $\hat{F}$ is also Hermitian.
With this definition, as in the single-valued case, criteria (\ref{eq:conv_criteria_1})
is fulfilled for matrix-valued functions.

\subsection{Semi-classical limit with matrix-valued functions}

In the following, we assume that the Fourier coefficients $F_{ab,n}(q)$
of $F$ are one times differentiable with respect to $q.$ This always
can be achieved for a regularization, by interpolation the points
$q(n,n)$ with a one times differentiable function. Analogously to
(\ref{eq:def_regularization}), we assume that the $F_{ab,n}(q)$
are again $0$ for $\left|n\right|\geq\delta$, where $\delta$ is
a constant that is small compared to $N$.

As in the case of the immersed cylinder (see (\ref{eq:ev_immersed_cylinder})),
we now can evaluate the matrix product of $\hat{F}$ with a further
matrix $\hat{G}$, which is a matrix regularization of a matrix-valued
function $G_{ab}(q,\varphi)$.
\[
(\hat{F}\hat{G})_{ab,nm}=\sum_{\begin{array}{c}
1\leq c\leq S\\
1\leq p\leq N
\end{array}}F_{ac,np}G_{cb,pm}=\sum_{\begin{array}{c}
1\leq c\leq S\\
1\leq p\leq N
\end{array}}F_{ac,p-n}\left(q(n,p)\right)G_{cb,m-p}\left(q(p,m)\right)
\]
\[
=\sum_{\begin{array}{c}
1\leq c\leq S\\
\left|n-p\right|<\delta\\
\left|p-m\right|<\delta
\end{array}}F_{ac,p-n}\left(q(n,m)+\frac{\beta}{N}(p-m)\right)G_{bc,m-p}\left(q(n,m)+\frac{\beta}{N}(p-n)\right)
\]
\[
=\sum_{\begin{array}{c}
1\leq c\leq S\\
\left|n-p\right|<\delta\\
\left|p-m\right|<\delta
\end{array}}F_{ac,p-n}G_{cb,m-p}+\frac{\beta}{N}\frac{\partial F_{ac,p-n}}{\partial q}\left(p-m\right)G_{cb,m-p}\;\;\;\;\;\;
\]
\[
\;\;\;\;\;\;+\frac{\beta}{N}F_{ac,p-n}\left(p-n\right)\frac{\partial G_{cb,m-p}}{\partial q}+\mathcal{O}\left(\frac{1}{N^{2}}\right)
\]
\begin{equation}
=Q\left(\sum_{1\leq c\leq S}F_{ac}G_{cb}+\frac{i\beta}{N}\left(\frac{\partial F_{ac}}{\partial q}\frac{\partial G_{cb}}{\partial\varphi}-\frac{\partial F_{ac}}{\partial\varphi}\frac{\partial G_{cb}}{\partial q}\right)\right)+\mathcal{O}\left(\frac{1}{N^{2}}\right)\label{eq:product_and_regularization}
\end{equation}
where in the second but last two step all functions are evaluated
at $q=q(n,m)$ and $Q$ is the regularization mapping, i.e. $\hat{F}=Q(F)$.
As in the immersed cylinder case, for solely continuous functions,
the result is $\hat{F}\hat{G}=Q(FG)+\mathcal{O}\left(\frac{1}{N}\right)$.
When the Fourier coefficient functions are not $0$ at $q_{1}$ and
$q_{2}$, then the considerations apply again, which have been made
with respect to the immersed cylinder relating to convergence in border.

From (\ref{eq:product_and_regularization}) it follows that also (\ref{eq:conv_criteria_2})
is fulfilled. However, in the case of single-valued functions, from
(\ref{eq:conv_criteria_2}) it is a trivial result that the commutator
of two regularized functions converges to $0$ for large $N$, i.e.
\begin{equation}
\lim_{N\rightarrow\infty}\left\Vert \left[Q_{N}(f),Q_{N}(g)\right]\right\Vert =0\label{eq:conv_creteria_4}
\end{equation}
since the functions commute. This is not longer the case for matrix-valued
functions. However, when we want that the regularized matrices converge
to a space of functions on a manifold, its is neccessary that (\ref{eq:conv_creteria_4})
is fulfilled. According to (\ref{eq:product_and_regularization})
this is only the case, when the commutator of the two matrix-valued
functions vanishes for $N$ going to infinity, i.e. $[F(q,\varphi),G(q,\varphi)]=\mathcal{O}\left(\frac{1}{N}\right)$.
If one finds $d$ matrix-valued functions $X^{\alpha}(q,\varphi)$,
which have the property that 
\begin{equation}
[X^{\alpha}(q,\varphi),X^{\beta}(q,\varphi)]=\mathcal{O}\left(\frac{1}{N}\right)\label{eq:nearly_commuting_coordinate_matrices}
\end{equation}
then the $N\times S$-dimensional matrices $Q_{N}(X^{\alpha})$ defined
by (\ref{eq:matrix_reg}), converge to a function space. When the
commutator (\ref{eq:nearly_commuting_coordinate_matrices}) vanishes
completely, then even a property analogous to (\ref{eq:conv_criteria_3})
is present.

(\ref{eq:matrix_reg}) is really a generalization of (\ref{eq:diagonal_matrix-valued function}),
since it suffices that the matrix-valued functions (\ref{eq:nearly_commuting_coordinate_matrices})
nearly commute. Since nearly commuting matrices can be nearly diagonalized,
one sees that there can be contributions to the $\mathcal{O}\left(\frac{1}{N}\right)$-part
of (\ref{eq:product_and_regularization}), which are not based on
a Poisson bracket.

In the case of commuting matrix-valued coordinate functions, these
can be diagonalized with a coordinate dependent unitary transformation
$U(q,\varphi)$. However, in general, $Q\left(U(q,\varphi)X^{\alpha}(q,\varphi)U(q,\varphi)^{\dagger}\right)\neq\hat{U}\hat{F}\hat{U}^{\dagger}$
and it is not obvious that the fuzzy space based on the original matrix-valued
functions is equivalent to the fuzzy space based on the diagonalized
matrix-valued functions.

\subsection{One diagonalized matrix}

As in the case of the immersed cylinder (\ref{eq:immersed_fuzzy_cylinder}),
we can consider the case, where one of the regularized coordinate
matrices, such as $\hat{X}^{d}=\hat{Z}$ is diagonal. This always
can be achieved by a unitary transformation on the fuzzy space, i.e.
after regularization. From (\ref{eq:matrix_reg}) it can be seen that
when the regularized matrix $\hat{Z}$ is diagonal, then the corresponding
matrix-valued function $Z(q,\varphi)$ has to be diagonal and independent
of $\varphi$. i.e. $Z_{ab}(q,\varphi)=$$z_{a}(q)\delta_{ab}$. The
commutator with a general matrix-valued function $F(q,\varphi)$ is
\begin{equation}
[Z(q),F(q,\varphi)]_{ab}=\left(z_{a}(q)-z_{b}(q)\right)F_{ab}(q,\varphi)
\end{equation}
When the differences of the diagonal entries of the function $Z(q)$
are of order $\frac{1}{N}$, the commutator of $Z(q)$ with every
other matrix-valued function $F(q,\varphi)$ is of the same order.
Thus, a fuzzy space, which is defined by matrix-valued functions $X^{\alpha}(q,\varphi)$
has a classical limit, when the diagonal matrix-valued function$Z(q)$
has entries, which differ only by order $\frac{1}{N}$, and when the
commutators of the other $X^{\alpha}(q,\varphi)$ are also of this
order.

An example with matrices of size $2\text{\ensuremath{\times}2}$,
which will become important in the following is
\begin{eqnarray}
Z & = & \left(\begin{array}{cc}
q\\
 & q+\frac{\beta(q)}{N}
\end{array}\right)\label{eq:off-diagonal example}\\
X^{\alpha} & = & \left(\begin{array}{cc}
 & x^{\alpha}(q,\varphi)\\
\bar{x}^{\alpha}(q,\varphi)
\end{array}\right)=\left(\begin{array}{cc}
 & r^{\alpha}(q,\varphi)e^{i\psi(q,\varphi)}\\
r^{\alpha}(q,\varphi)e^{-i\psi(q,\varphi)}
\end{array}\right)\nonumber 
\end{eqnarray}
where $\beta(q)$ is a function of order 1 and the $r^{\alpha}$ and
$\psi$ are real-valued functions. For example, two interlaced mirror
symmetric fuzzy spaces (\ref{eq:def_mirror_cyl_interlaced}), such
as two interlaced cylinders, have the form of such an ``off-diagonal
fuzzy space''. In the next section we will show that this is also
the case for the circle-to-eight space (\ref{eq:circle-to-eight}),
when it is considered as a regularization based on $2\times2$-matrix
valued functions.

The $X^{\alpha}(q,\varphi)$ of (\ref{eq:off-diagonal example}) commute
with each other, since
\begin{equation}
[\left(\begin{array}{cc}
 & f\\
\bar{f}
\end{array}\right),\left(\begin{array}{cc}
 & g\\
\bar{g}
\end{array}\right)]=\left(\begin{array}{cc}
f\bar{g}-g\bar{f}\\
 & -(f\bar{g}-g\bar{f})
\end{array}\right)
\end{equation}
The right-hand side vanishes, when $f\bar{g}=g\bar{f}$, which is
the case for the $x^{\alpha}$ defined in (\ref{eq:off-diagonal example}).

However, the matrix-valued functions $Z(q)$ and $X^{\alpha}(q,\varphi)$
of (\ref{eq:off-diagonal example}) are in general not simultaneously
diagonalizable. The corresponding matrix regularization does not split
into a direct sum of immersed cylinders. On the other hand, in the
limit $N\rightarrow\infty$ the functions $Z(q)$ and $X^{\alpha}(q,\varphi)$
are commuting matrices, which can be diagonalized simultaneously and
therefore can be seen as the direct sum of two submanifolds.

\subsection{Immersed cylinder based on matrix-valued functions}

We will now show that the immersed fuzzy cylinder as defined in (\ref{eq:fuzzy_cylinder})
can be seen as fuzzy space defined by $2\times2$-matrix-valued functions. 

In general let $f(q,\varphi)$ be a function and let $\hat{f}$ be
its (single-valued) matrix regularization, i.e. $\hat{f}_{ij}(q)=f_{j-i}(q(i,j))$.
We can now ask the question, whether there is a matrix regularization
of a matrix-valued function $F(q,\varphi)$, which results in the
the same matrix $\hat{f}$. To construct such a function $F(q,\varphi)$,
we can decompose the matrix $\hat{f}$ into $2\times2$ cells $\hat{F}_{ij}$
\begin{equation}
\hat{F}_{ij}\mathrm{=\left(\begin{array}{cc}
f_{2(j-i)}\left(q(2i,2j)\right) & f_{2(j-i)+1}\left(q(2i,2j+1)\right)\\
f_{2(j-i)-1}\left(q(2i+1,2j)\right) & f_{2(j-i)}\left(q(2i+1,2j+1)\right)
\end{array}\right)}
\end{equation}
and compare this with the Fourier series of the matrix-valued function
$F(q,\varphi)=\sum_{n}F_{n}e^{in\varphi}$. For $q(i,j)=\frac{i+j}{2N}$
we see that the matrix $\hat{f}$ also is a regularization of 
\begin{equation}
F(q,\varphi)=\mathrm{\left(\begin{array}{cc}
f_{2n}\left(2q\right) & f_{2n+1}\left(2q+\frac{1}{2N}\right)\\
f_{2n-1}\left(2q+\frac{1}{2N}\right) & f_{2n}\left(2q+\frac{1}{N}\right)
\end{array}\right)e^{in\varphi}}
\end{equation}
This results in
\[
F(q,\varphi)=
\]
 
\begin{equation}
\frac{1}{2}\left(\begin{array}{cc}
f\left(2q,\frac{\varphi}{2}\right)+f\left(2q,\frac{\varphi+2\pi}{2}\right) & \left(f\left(2q+\frac{1}{2N},\frac{\varphi}{2}\right)-f\left(2q+\frac{1}{2N},\frac{\varphi+2\pi}{2}\right)\right)e^{-\frac{i\varphi}{2}}\\
\left(f\left(2q+\frac{1}{2N},\frac{\varphi}{2}\right)-f\left(q+\frac{1}{2N},\frac{\varphi+2\pi}{2}\right)\right)e^{\frac{i\varphi}{2}} & f\left(2q+\frac{1}{N},\frac{\varphi}{2}\right)+f\left(2q+\frac{1}{N},\frac{\varphi+2\pi}{2}\right)
\end{array}\right)\label{eq:F_immersed_cylinder}
\end{equation}
i.e. the diagonal elements are the double periodic part of $f$ and
the off-diagonal elements are the phase-shifted double anti-periodic
part of $f$.

For coordinate functions $x^{\alpha}(q,\varphi)$, which are double
anti-periodic, i.e. $x^{\alpha}\left(q,\frac{\varphi}{2}+\pi\right)+x^{\alpha}\left(q,\frac{\varphi}{2}\right)=0$,
(such as those for the circle-to-eight manifold (\ref{eq:circle-to-eight})),
the corresponding matrix-valued coordinate functions (\ref{eq:F_immersed_cylinder})
reduce to
\begin{equation}
X^{\alpha}(q,\varphi)=\left(\begin{array}{cc}
 & x^{\alpha}\left(2q+\frac{1}{2N},\frac{\varphi}{2}\right)e^{-\frac{i\varphi}{2}}\\
x^{\alpha}\left(2q+\frac{1}{2N},\frac{\varphi}{2}\right)e^{\frac{i\varphi}{2}}
\end{array}\right)\label{eq:F_double_anti_periodic}
\end{equation}
i.e. $X^{\alpha}(q,\varphi)$ have the form (\ref{eq:off-diagonal example})
with $\psi(q,\varphi)=-\frac{i\varphi}{2}$.

For a function $z(q)$, which does not depend on $\varphi$, the result
is 
\begin{equation}
Z(q)=\left(\begin{array}{cc}
z(2q)\\
 & z(2q+\frac{1}{N})
\end{array}\right)
\end{equation}
The function $Z(q)$ has the property that the difference of its entries
is of order $\frac{1}{N}$. Thus, the matrix-valued functions $X^{\alpha}(q,\varphi)$
and $Z(q)$ are an ``off-diagonal fuzzy space'' (\ref{eq:off-diagonal example})
as described in the previous section.

Note that all matrix-valued functions $X^{\alpha}(q,\varphi)$ and
$Z(q)$ depend on $N$. However in the large $N$ limit, the term
$\frac{1}{N}$ can be set to zero and the matrix regularization of
the $X^{\alpha}(q,\varphi)$ and $Z(q)$ splits into a direct sum.
The functions $X^{\alpha}(q,\varphi)$ and $Z(q)$ can be seen as
the interlaced form of two fuzzy spaces. As is shown below, see (\ref{eq:interlacing}),
the phase factors $e^{\pm\frac{i\varphi}{2}}$ do not contribute to
the classical limit. Thus, a fuzzy space defined as a regularization
of double anti-periodic coordinate functions $x^{\alpha}(q,\varphi)$
and $z(q)$ in the large $N$ limit can be seen as the direct sum
of two fuzzy spaces defined by $\pm x^{\alpha}\left(2q,\frac{\varphi}{2}\right)$
and $z(2q)$.

\subsection{Non-constant unitary transformations }

We already have considered the case of a constant unitary transformation
of a matrix-valued function. Such a constant unitary transformation
(\ref{eq:constant_unitary}) corresponds to a unitary transformation
in the matrix regularization, which is the $N$-fold direct product
of this constant unitary transformation with itself. One can know
examine the relationship between unitary transformations$U(q,\varphi)$
applied on the side of the matrix-valued functions, i.e. before regularization,
and unitary transformations on the side of the fuzzy space, i.e. after
regularization.

In general, when the unitary transformation$U(q,\varphi)$ is coordinate
dependent, regularization (\ref{eq:matrix_reg}) and multiplication
do not commute. In particular, when $\hat{U}=Q_{N}(U)$ is the matrix
regularization of $U(q,\varphi)$ $\hat{U}$ need not be unitary and
$\hat{U}^{\dagger}$need not be its inverse.

\subsubsection{$q$-dependent unitary transformation}

In the case, when $U(q)$ depends only on $q$, then it follows from
(\ref{eq:matrix_reg}) that $Q_{N}(U)^{\dagger}=Q_{N}(U^{\dagger})$
is the inverse of $Q_{N}(U).$ At least in this case, in the large
$N$ limit, the manifold defined by the matrix-valued coordinate functions,
when present, stays the same, since
\begin{equation}
Q_{N}(U)Q_{N}(F)Q_{N}(U^{\dagger})=Q_{N}(UFU^{\dagger})+\mathcal{O}\left(\frac{1}{N}\right)
\end{equation}
according to (\ref{eq:product_and_regularization}). For example,
the pure $q$-dependent diagonal unitary transformation 
\begin{equation}
U(q)=\frac{1}{\sqrt{2}}\left(\begin{array}{cc}
e^{i\psi_{1}(q)}\\
 & e^{i\psi_{1}(q)}
\end{array}\right)
\end{equation}
applied to an off-diagonal matrix-valued function $F(q,\varphi)$,
such as the $X^{\alpha}(q,\varphi)$ in (\ref{eq:off-diagonal example}),
results in a further phase shift of the off-diagonal elements

\begin{equation}
U(q)\left(\begin{array}{cc}
 & f(q,\varphi)\\
\overline{f(q,\varphi)}
\end{array}\right)U(q)=\left(\begin{array}{cc}
 & e^{i(\left(\psi_{1}(q)-\psi_{2}(q)\right)}f(q,\varphi)\\
e^{-i(\left(\psi_{1}(q)-\psi_{2}(q)\right)}\overline{f(q,\varphi)}
\end{array}\right)\label{eq:phase_shifting}
\end{equation}
Thus, the phase of (\ref{eq:off-diagonal example}), when only $q$-dependent
as in (\ref{eq:F_double_anti_periodic}) can be removed with a suitable
unitary transformation in the large $N$ limit. In particular, The
matrix-valued function 
\begin{equation}
U(q)=\frac{1}{\sqrt{2}}\left(\begin{array}{cc}
1 & -e^{i\psi(q)}\\
e^{-i\psi(q)} & 1
\end{array}\right)
\end{equation}
applied to the $X^{\alpha}(q,\varphi)$ of (\ref{eq:off-diagonal example})
with only $q$-dependent $\psi(q)$ results in 
\begin{equation}
U(q)X^{\alpha}(q,\varphi)U^{\dagger}(q)=\left(\begin{array}{cc}
-r^{\alpha}(q,\varphi)\\
0 & r^{\alpha}(q,\varphi)
\end{array}\right)\label{eq:interlacing}
\end{equation}
and leaves $Z(q)$ invariant. Only the magnitudes $r^{\alpha}(q,\varphi)$
of the functions $x^{\alpha}(q,\varphi)$ define the classical limit
of the fuzzy space. The phase can be seen as pure non-commutative
artifact.

\subsubsection{Block transformations}

A further interesting example arises in the application of unitary
block matrices $\hat{U}$, which are based on constant unitary matrices
that are applied only to some of the matrix blocks of the regularized
matrices, such as 

\begin{equation}
\hat{U}=\sum_{n>n_{o}}U\hat{e}_{nn}\label{eq:block_transformation}
\end{equation}
where $n_{0}$ is some number between $1$ and $N$ and $U$ is a
unitary $S\times S$-matrix. This unitary transformation $\hat{U}$
can be seen as the regularization of the matrix-valued function $U(q)=\theta(q)U$
where $\theta$ is a step function, which is $0$ below a $q_{0}=q(n_{0},n_{0})$
($q$ is the discretizing function) and which is $1$ above.

For example, let $\hat{F}$ be the regularization of the matrix-valued
function $F(q,\varphi)=\sum_{n}F_{n}(q)e^{in\varphi}$ , i.e. 
\begin{equation}
\hat{F}=\left(\begin{array}{cccccc}
\ddots & \ddots & \ddots & \ddots & \ddots & \ddots\\
\ddots & F_{o} & F_{1} & F_{2} & F_{3} & \ddots\\
\ddots & F_{-1} & F_{o} & F_{1} & F_{2} & \ddots\\
\ddots & F_{-2} & F_{-1} & F_{0} & F_{1} & \ddots\\
\ddots & F_{-3} & F_{2} & F_{-1} & F_{0} & \ddots\\
\ddots & \ddots & \ddots & \ddots & \ddots & \ddots
\end{array}\right)\ \hat{U}=\left(\begin{array}{ccc|ccc}
\ddots &  & \\
 & 1 & \\
 &  & 1\\
\hline  &  &  & U\\
 &  &  &  & U\\
 &  &  &  &  & \ddots
\end{array}\right)
\end{equation}
where we have omitted the arguments of the $S\times S$-matrices$F_{i}$.
Then
\begin{equation}
\hat{U}^{\dagger}\hat{F}\hat{U}=\left(\begin{array}{ccc|ccc}
\ddots & \ddots & \ddots & \ddots & \ddots & \ddots\\
\ddots & F & F_{1} & F_{2}U & F_{3}U & \ddots\\
\ddots & F_{-1} & F_{o} & F_{1}U & F_{2}U & \ddots\\
\hline \ddots & U^{\dagger}F_{-2} & U^{\dagger}F_{-1} & U^{\dagger}F_{0}U & U^{\dagger}F_{1}U & \ddots\\
\ddots & U^{\dagger}F_{-3} & U^{\dagger}F_{2} & U^{\dagger}F_{-1}U & U^{\dagger}F_{0}U & \ddots\\
\ddots & \ddots & \ddots & \ddots & \ddots & \ddots
\end{array}\right)\label{eq:unitary_block}
\end{equation}
One sees that the upper left part is not transformed at all, that
the lower right part is completely transformed with $U$, and that
the off-diagonal parts are only transformed on one side. Later we
will encounter such a transformation in an example.

\section{Interpolation of fuzzy spaces}

We have seen that the matrix-valued functions, which define both the
circle-to-eight fuzzy space and the interlaced cylinder fuzzy space,
have the form (\ref{eq:off-diagonal example}), i.e. the $Z(q)$-coordinate
function is diagonal and the $X^{\alpha}(q,\varphi)$-coordinate functions
are off-diagonal. In the following we will show that it is possible
to interpolate the corresponding matrices along the $q$-coordinate
such that also the part, which depends on the coordinate functions
of both spaces has the form (\ref{eq:off-diagonal example}). In this
way, we arrive at a fuzzy space that has a defined classical limit
everywhere along the $q$-coordinate.

Based on this, we explicitly construct a fuzzy space that in the classical
limit corresponds to a string vertex, i.e. a submanifold that has
three ends formed like cylinders, which are interconnected by a branching.

\subsection{Extension of circle-to-eight by two cylinders (classical case)}

In this section, we construct a classical string vertex from two coordinate
patches. We hope that this will render the construction based on matrix-valued
functions in the next section more clear. 

In general, we have two manifolds that overlap in a middle part between
$q_{2}$ and $q_{3}$ along the $q$-coordinate. The first manifold
(the circle-to-eight manifold) is defined between $q_{1}$ and $q_{3}$
and the second manifold (the two cylinders) are defined between $q_{2}$
and $q_{4}$. 

As described above, the circle-to-eight (\ref{eq:circle-to-eight})
is defined by

\begin{equation}
x(q,\varphi)=r(q,\varphi)\cos\varphi\ \ \ y(q,\varphi)=r(q,\varphi)\sin\varphi\ \ \ z(q,\varphi)=q
\end{equation}
with
\begin{equation}
r(q,\varphi)=r_{1}(q)+r_{2}(q)\cos2\varphi
\end{equation}
such that $r_{2}(q_{1})=0$ and $r_{1}(q_{2})=r_{2}(q_{2})=r_{1}$
(see (\ref{eq:circle-to-eight}) above). We assume that the functions
$r_{1}$ and $r_{2}$ stay constant between $q_{2}$ and $q_{3}$.
The functions $x$ and $y$ are double anti-periodic, i.e. 
\begin{equation}
x(q,\frac{\varphi}{2})=-x(q,\frac{\varphi}{2}+\pi)\ \ \ y(q,\frac{\varphi}{2})=-y(q,\frac{\varphi}{2}+\pi)
\end{equation}
This is due to the fact that the submanifold defined by $x$, $y$
and $z$ is axis-symmetric to the $z$-axis. The parts of the submanifolds
for $\varphi\in[\frac{\pi}{2},\frac{3\pi}{2}]$ and $\varphi\in[\frac{3\pi}{2}\sim-\frac{\pi}{2},\frac{5\pi}{2}\sim\frac{\pi}{2}]$
are symmetric. Between $q_{2}$ and $q_{3}$, the ``crossing'' of
the eight is located at $\varphi=\frac{\pi}{2}$ and $\varphi=-\frac{\pi}{2}$. 

We can define two other coordinate patches for the circle-to-eight
manifold. In particular, the right half with $\varphi\in[-\frac{\pi}{2},\frac{\pi}{2}]$
is defined via $x_{R}(q,\varphi)=x(q,\frac{\varphi}{2}+\frac{\pi}{2})$
and $y_{R}$ is defined respectively. For the left half, the function
$y_{L}=y_{R}$ is the same and the function $x_{L}=-x_{R}$ is the
negative of the one of the right half. Between $q_{2}$ and $q_{3}$,
the Fourier coefficients $x_{Rm}$, $y_{Rm}$, $x_{Lm}$ and $y_{Rm}$
of these functions for $m<-1$ and $m>1$ can be faded out and the
Fourier coefficients for $m=-1,0,1$ can be continuously adjusted
such they go over to the ones of two cylinders as in (\ref{eq:two_cylinders})
, which are defined between $q_{3}$ and $q_{4}.$

More general, let $x_{1}^{\alpha}$ be the coordinate functions of
one of the halves of the first manifold between $q_{1}$ and $q_{3}$
defined by double anti-periodic functions and let $x_{2}^{\alpha}$
be the coordinate functions for a cylinder between $q_{2}$ and $q_{4}$.
Then, the middle part between $q_{2}$ and $q_{3}$ can be defined
by interpolation

\[
x^{\alpha}(q,\varphi)=\theta_{1}(q)x_{1}^{\text{\ensuremath{\alpha}}}(q,\frac{\varphi}{2}+\frac{\pi}{2})+\theta_{2}(q)x_{2}^{\alpha}(q,\varphi)
\]
 where $\theta_{1}$ and $\theta_{2}$ are two continuous functions
with $\theta_{1}(q<q_{2})=1$, $\theta_{1}(q>q_{3})=0$, $\theta_{2}(q<q_{2})=0$
and $\theta_{2}(q>q_{3})=1$. Correspondingly, the other half can
be defined. When appropriately choosing the parameters for $x_{2}^{\alpha}$,
i.e. the part with the two cylinders, the resulting manifold will
have the form of a string vertex.

Even more general, we can include a $q$-dependent coordinate transformation
in the interpolation. 
\begin{equation}
x^{\alpha}(q,\varphi)=\theta_{1}(q)x_{1}^{\text{\ensuremath{\alpha}}}\left(q,\frac{\varphi}{2}+\frac{\pi}{2}+\pi\beta(q)\right)+\theta_{2}(q)x_{2}^{\alpha}\left(q,\varphi+\pi\beta(q)\right)\label{eq:cont_coord_trafo_classical}
\end{equation}
with a continous function $\beta$ that is equal to $-\frac{1}{2}$
in the part of the space with the circle-to-eight, i.e. for $q<q_{2}$
and that is $0$ in the pure double cylinder part, i.e. for $q>q_{3}$.
The function $\beta$ has the advantage that there is no need for
applying a coordinate transformation on the parts outside of the transition
part between $q_{2}$ and $q_{3}$ and the Fourier coefficients of
$x_{1}^{\alpha}$ and $x_{2}^{\alpha}$ outside of the transition
part need not be adapted.

We will carry over this construction in the following to fuzzy spaces
of the form (\ref{eq:off-diagonal example}), where $Z(q)$ is diagonal
and the $X^{\alpha}(q,\varphi)$ are off-diagonal.

\subsection{Interpolating between off-diagonal fuzzy spaces}

As just mentioned, we start with two fuzzy spaces as defined in (\ref{eq:off-diagonal example})
. We assume that the first fuzzy space has off-diagonal functions
$x^{\alpha}=f_{1}^{\alpha}e^{-i\frac{\varphi}{2}}$, which are defined
for $q\in[q_{1},q_{3}]$, and that the second fuzzy space has off-diagonal
functions $x_{2}^{\alpha}=f_{2}^{\alpha}$, which are defined for
$q\in[q_{2},q_{4}],$where $q_{2}<q_{3}$. The first fuzzy space is
of the form (\ref{eq:F_double_anti_periodic}), such as the circle-to-eight
space, and the second space is of the form (\ref{eq:def_mirror_cyl_interlaced})
such as two interlaced cylinders.

In correspondence with (\ref{eq:cont_coord_trafo_classical}), an
interpolation between the off-diagonal functions $x_{1}^{\alpha}$
and $x_{2}^{\alpha}$ is introduced by 
\begin{equation}
x^{\alpha}(q,\varphi)=\left(\theta_{1}(q)f_{1}^{\alpha}(q,\frac{\varphi}{2}+\frac{\pi}{2}+\pi\beta(q))+\theta_{2}(q)f_{2}^{\alpha}(q,\varphi+\pi\beta(q))\right)e^{i\alpha(q)\varphi+i\gamma(q)}\label{eq:interpolation}
\end{equation}
with the functions $\theta_{1}$, $\theta_{2}$ and $\beta$ as in
the previous section. The continuous function $\alpha$ has the same
properties as $\beta$ and is used for a continuous transition of
the phase factor. The continuous function $\gamma$ will be used for
adapting a final phase shift of the the function $x^{\alpha}$, which
does does not change the classical limit (see (\ref{eq:phase_shifting})
). 

With such definitions, the resulting space defined by 
\begin{equation}
X^{\alpha}=\left(\begin{array}{cc}
 & x^{\alpha}\\
\bar{x}^{\alpha}
\end{array}\right)\ \ Z=\theta_{1}Z_{1}+\theta_{2}Z_{2}\label{eq:all_matrices_interpolated}
\end{equation}
 is equal to the first space for $q<q_{2}$ and equal to the second
space for $q>q_{3}$. Importantly, between $q_{2}$ and $q_{3}$,
the matrix-valued functions $X^{\alpha}$ and $Z$ are also of the
form (\ref{eq:off-diagonal example}), i.e. their commutators vanish
to first order and thus the regularization (\ref{eq:matrix_reg})
has a classical limit. Due to our considerations above, we know that
the manifold of the classical limit is defined by the magnitude of
the functions $x^{\alpha}$, i.e. by (\ref{eq:cont_coord_trafo_classical}).

After regularization (\ref{eq:matrix_reg}), the part of the coordinate
matrices $\hat{X}^{\alpha}$ belonging to $q<q_{2}$ (the left upper
part) looks like the two left diagrams of Fig. 2. For interlaced cylinders,
the part belonging to $q>q_{3}$ (the lower right part) looks like
the two lower left diagrams of Fig. 3. With the interpolation (\ref{eq:interpolation}),
the part in between is filled out. One sees that due to the interlacing
of the second space, the rows of Fourier coefficients of the first
space can go over to the rows of Fourier coefficients of the second
space in a smooth way. Due to the interpolation, a bifurcation of
the Fourier coefficient functions of the first space into the double
periodic Fourier coefficient functions of the second space takes place.

We now determine the Fourier coefficients of the middle part of the
coordinate matrices $\hat{X}^{\alpha}$ with $q_{2}<q<q_{3}$. In
the following we will omit the label $\alpha$ and will concentrate
on one of the coordinates. With 
\begin{eqnarray}
f_{1}(q,\frac{\varphi}{2})e^{-\frac{i\varphi}{2}} & = & \sum_{n}f_{1,n}(q)e^{in\varphi}\\
f_{2}(q,\varphi) & = & \sum_{n}f_{2,n}(q)e^{in\varphi}
\end{eqnarray}
where for the first fuzzy space, the Fourier coefficients $f_{1,n}$
of the respective immersed fuzzy cylinder (\ref{eq:fourier_single_valued})
have been used, the Fourier coefficients of the interpolating functions
become
\begin{eqnarray}
f_{m}(q) & = & e^{i\gamma(q)}\sum_{n}\left(\theta_{1}(q)f_{1,n}(q)\intop_{0}^{2\pi}\frac{d\varphi}{2\pi}e^{i\left(n-m+\frac{1}{2}+\alpha(q)\right)\varphi+i\pi\left(\frac{1}{2}+\beta(q)\right)n}\right.\nonumber \\
 &  & \left.\;\;\;\;\;\;\;\;\;\;\;\;\;\;\;\;\;\;\;\;+\theta_{2}(q)f_{2,n}(q)\intop_{0}^{2\pi}\frac{d\varphi}{2\pi}e^{i\left(n-m+\alpha(q)\right)\varphi+i\pi\beta(q)n}\right)\nonumber \\
 & = & e^{i\alpha(q)\pi+i\gamma(q)}\frac{1}{\pi}\sum_{n}\left(\theta_{1}(q)\sin\pi\left(\frac{1}{2}+\alpha(q)\right)\frac{f_{1,n}e^{i\pi\left(\frac{1}{2}+\beta(q)\right)n}}{n-m+\frac{1}{2}+\alpha(q)}\right.\nonumber \\
 &  & \left.\;\;\;\;\;\;\;\;\;\;\;\;\;\;\;\;\;\;\;\;+\theta_{2}(q)\sin\pi\alpha(q)\frac{f_{2,n}e^{i\pi\beta(q)n}}{n-m+\alpha(q)}\right)
\end{eqnarray}
As (like for ordinary Fourier series)

\begin{equation}
\lim_{\alpha\rightarrow-\frac{1}{2}}\frac{1}{\pi}\frac{f_{1,n}}{n-m+\frac{1}{2}+\alpha}\sin\pi\left(\frac{1}{2}+\alpha(q)\right)=f_{1,n}\delta_{nm}
\end{equation}

\begin{equation}
\lim_{\alpha\rightarrow0}\frac{1}{\pi}\frac{f_{2,n}}{n-m+\alpha}\sin\pi\alpha=f_{2,n}\delta_{nm}
\end{equation}
at the the points $q=q_{2}$ and $q=q_{3}$, the Fourier coefficients
$f_{m}$ converge to the respective ones of the first and second fuzzy
space, when one chooses $\lambda=-\pi\alpha$ at these points. We
will choose $\lambda=-\pi\alpha$ for all $q$.

Furthermore, we are free to choose $\beta=\alpha$, which results
in
\begin{equation}
f_{m}(q)=\frac{1}{\pi}\sum_{n}\left(\theta_{1}(q)\cos\pi\alpha(q)\frac{f_{1,n}e^{i\pi\left(\frac{1}{2}+\alpha(q)\right)n}}{n-m+\frac{1}{2}+\alpha(q)}+\theta_{2}(q)\sin\pi\alpha(q)\frac{f_{2,n}e^{i\pi\alpha(q)n}}{n-m+\alpha(q)}\right)\label{eq:Fourier_coefficient_interpolated}
\end{equation}

When $f_{1,n}(q)=0$ and $f_{2,n}(q)=0$ for $\left|n\right|>\delta$,
which we have assumed for (\ref{eq:product_and_regularization}),
we can estimate for $m>\delta$ 
\begin{eqnarray}
\left|f_{m}(q)\right| & \leq & \frac{1}{\pi}\sum_{\left|n\right|\leq\delta}\left(\left(\frac{f_{1,n}}{n-m+\frac{1}{2}+\alpha(q)}\right)^{2}+\left(\frac{f_{2,n}}{n-m+\alpha(q)}\right)^{2}\right)\nonumber \\
 & \leq & C\sum_{\left|n\right|\leq\delta}\left(\frac{1}{n-m}\right)^{2}<\frac{C(2\delta+1)}{m^{2}}\label{eq:estimation}
\end{eqnarray}
with $C=\frac{2}{\pi}\max_{\left|n\right|\leq\delta}\{f_{1,n},f_{2,n}\}$.
We see that in the interpolation part, the Fourier coefficients $f_{m}(q)$
for raising $m$ go at least quadratically to $0$. This was to be
expected, since the interpolated function $x^{\alpha}$ (\ref{eq:interpolation})
has the same properties (such as differentiability, etc.) as the functions
$x_{1}^{\alpha}$ and $x_{2}^{\alpha}$ being interpolated, when the
functions $\theta_{1}$, $\theta_{2}$ and $\alpha$ are sufficiently
smooth. Therefore, it is possible to truncate the Fourier series $f_{m}(q)$
at a specific $m<\tilde{\delta}$, for a $\tilde{\delta}>\delta$,
which is also small compared to $N$. This $\tilde{\delta}$ can become
larger with growing $N$ approximating the manifold better and better
and (\ref{eq:product_and_regularization}) stays valid.

The Fourier coefficient $\bar{f}_{n}$ of the complex conjugate of
a function is in general the complex conjugate of the Fourier coefficient
for $-n$, i.e. $\bar{f}_{n}=\overline{f_{-n}}$. In the present case
$f_{1,n}=\tilde{f}_{2n+1}$ of a real valued function $\tilde{f}$
and therefore $\bar{f}_{1,n}=\overline{f_{1,-n}}=\overline{\tilde{f}_{-2n+1}}=\tilde{f}_{2n-1}=\tilde{f}_{2(n-1)+1}=f_{1,n-1}$.
In summary
\begin{equation}
\bar{f}_{m}(q)=-\frac{1}{\pi}\sum_{n}\left(\theta_{1}(q)\cos\pi\alpha(q)\frac{f_{1,-n-1}e^{i\pi\left(\frac{1}{2}+\alpha(q)\right)n}}{n-m-\frac{1}{2}-\alpha(q)}+\theta_{2}(q)\sin\pi\alpha(q)\frac{f_{2,n}e^{i\pi\alpha(q)n}}{n-m-\alpha(q)}\right)
\end{equation}
If we additionally choose

\begin{equation}
\theta_{1}(q)=-\lambda(q)\sin\pi\alpha(q)\ \;\;\theta_{2}(q)=\lambda(q)\cos\pi\alpha(q)
\end{equation}
a very compact form for the Fourier coefficients can be achieved.
These functions $\theta_{1}$ and $\theta_{2}$ fulfill the demanded
properties, if $\lambda(q)$ is a continuous function, which is equal
to $1$ outside of the interval $[q_{2},q_{3}]$ (For the ``-''
in front of the sine function note that $-\frac{1}{2}\leq\alpha\leq0$).
If we further demand that a fuzzy space interpolated with itself stays
the same, it must be that $\theta_{1}+\theta_{2}=1$ and $\lambda$
is 
\begin{equation}
\lambda(q)=\frac{1}{\cos\pi\alpha(q)-\sin\pi\alpha(q)}
\end{equation}
We arrive at

\begin{equation}
f_{m}(q)=\frac{1}{\pi}\frac{\cos\pi\alpha(q)\sin\pi\alpha(q)}{\cos\pi\alpha(q)-\sin\pi\alpha(q)}\sum_{n}\left(-\frac{f_{1,n}e^{i\pi\left(\frac{1}{2}+\alpha(q)\right)n}}{n-m+\frac{1}{2}+\alpha(q)}+\frac{f_{2,n}e^{i\pi\alpha(q)n}}{n-m+\alpha(q)}\right)
\end{equation}

\subsection{A fuzzy vertex with classical limit\label{sub:A-fuzzy-vertex}}

To give an explicit example, we numerically calculate the matrix regularization
of (\ref{eq:Fourier_coefficient_interpolated}) for the circle-to-eight
example interpolated with two cylinders. We define for the circle-to-eight
part (see (\ref{eq:circle-to-eight}) )

\begin{equation}
\tilde{x}_{1,m}=\frac{\frac{r_{2}}{4}e^{-2i\pi\left(\frac{1}{2}+\alpha\right)}}{-2-m+\frac{1}{2}+\alpha}+\frac{\frac{1}{2}(r_{1}+\frac{r_{2}}{2})e^{-i\pi\left(\frac{1}{2}+\alpha\right)}}{-1-m+\frac{1}{2}+\alpha}+\frac{\frac{1}{2}(r_{1}+\frac{r_{2}}{2})}{-m+\frac{1}{2}+\alpha}+\frac{\frac{r_{2}}{4}e^{i\pi\left(\frac{1}{2}+\alpha\right)}}{1-m+\frac{1}{2}+\alpha}
\end{equation}
\begin{equation}
\tilde{y}_{1,m}=\frac{-\frac{ir_{2}}{2}e^{-2i\pi\left(\frac{1}{2}+\alpha\right)}}{-2-m+\frac{1}{2}+\alpha}+\frac{-\frac{i}{2}(r_{1}-\frac{r_{2}}{2})e^{-i\pi\left(\frac{1}{2}+\alpha\right)}}{-1-m+\frac{1}{2}+\alpha}+\frac{\frac{i}{2}(r_{1}-\frac{r_{2}}{2})}{-m+\frac{1}{2}+\alpha}+\frac{\frac{ir_{2}}{2}e^{i\pi\left(\frac{1}{2}+\alpha\right)}}{1-m+\frac{1}{2}+\alpha}
\end{equation}
and for the interlaced cylinder part (see (\ref{eq:two_cyl_interlaced})
)

\begin{equation}
\tilde{x}_{2,m}=\frac{re^{-i\pi\alpha}}{-1-m+\alpha}+\frac{x_{0}(q)}{-m+\alpha}+\frac{re^{i\pi\alpha}}{1-m+\alpha}
\end{equation}
\begin{equation}
\tilde{y}_{2,m}=\frac{-ire^{-i\pi\alpha}}{-1-m+\alpha}+\frac{ire^{i\pi\alpha}}{1-m+\alpha}
\end{equation}
Then
\begin{equation}
x_{m}=\frac{1}{\pi}\left(\theta_{1}\tilde{x}_{1,m}+\theta_{2}\tilde{x}_{2,m}\right)
\end{equation}
\begin{equation}
y_{m}=\frac{1}{2}\left(\theta_{1}\tilde{y}_{1,m}+\theta_{2}\tilde{y}_{2,m}\right)
\end{equation}
With (\ref{eq:F_immersed_cylinder}) and (\ref{eq:all_matrices_interpolated}),
the $z$-coordinate becomes

\begin{equation}
Z(q,\varphi)=\mathrm{\left(\begin{array}{cc}
\theta_{1}2q+\theta_{2}q\\
 & \theta_{1}(2q+\frac{1}{N})+\theta_{2}q
\end{array}\right)}
\end{equation}
For the numerically example, shown in Fig. 4, we have chosen $r_{1}=1$,
$r=1$, $x_{o}(q)=0.7+0.3q$, $\alpha(q)=\frac{1}{2}\left(h(2q-3)-1\right)$,
$\theta_{2}(q)=h(2q-3)$ and $\theta_{1}(q)=1-\theta_{2}(q)$ with
the spline $h$ as defined in chapter 2. These parameters and functions
result in a manifold for the classical limit, which does not intersect
with itself (see the right, bottom picture of Fig. 4). The two upper
pictures and the left lower picture of Fig. 4 show a visualization
of $\hat{X}$, $\hat{Y}$ and $\hat{Z}$. These matrices have been
generated with the discretizing function $q(n,m)=4\frac{n+m}{2N}-1$
for $2N=60$. In the diagrams for $\hat{X}$, $\hat{Y}$ and $\hat{Z}$,
matrix entries with norm smaller than $0.1$ are depicted with a small
dot. 

It can be seen that the diagonal rows of matrices $\hat{X}$ and $\hat{Y}$
smoothly go over from the four rows of the circle-to-eight space into
the four double periodic rows of the two cylinders. In the interpolating
part, further diagonal rows are present, which, however, have rather
small norms, as expected from (\ref{eq:estimation}).

\begin{figure}
\includegraphics[scale=0.5]{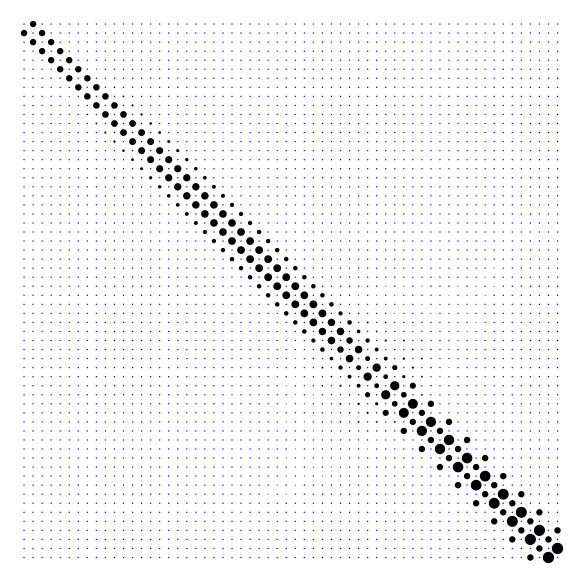}\includegraphics[scale=0.5]{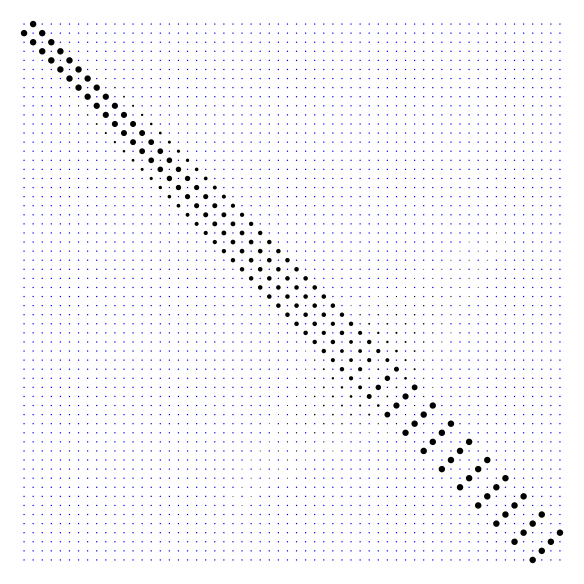}

\includegraphics[scale=0.5]{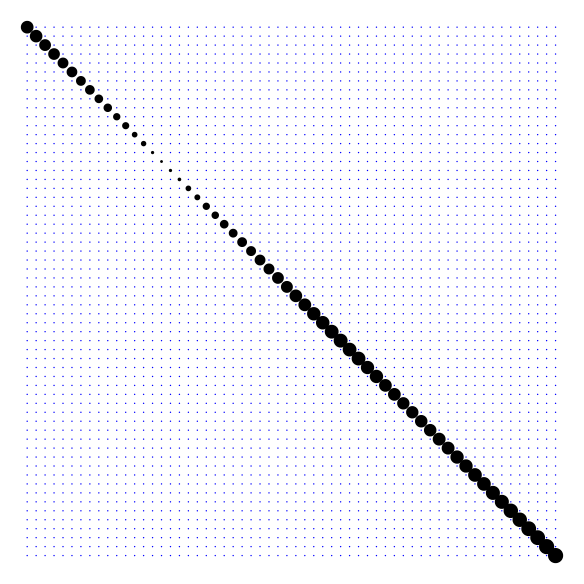} \includegraphics[scale=0.6]{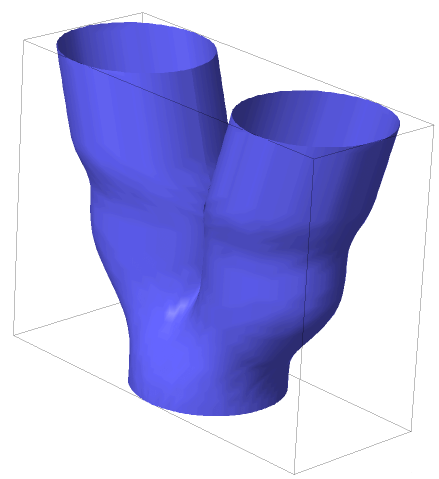}

\caption{Circle-to-eight extended by two cylinders}
\end{figure}

Since the ends of the string vertex are open, the matrix entries of
the commutators of the matrices for $N\rightarrow\infty$ only go
to $0$ within borders. Again, the ends of the string vertex can be
closed by letting all functions $f_{m}(q)$ go to $0$ at the ends.
This also can be done by multiplying the matrix-valued functions (\ref{eq:all_matrices_interpolated})
with a function, which goes continuously to $0$ at $q_{1}$ and $q_{4}$.

We also have calculated the commutators of the numerically determined
matrices for $N$ up to $120$ and verified that the maximal values
of the norm of the matrix entries of these commutators between borders,
i.e. for $5<n<N-5$ decreases with raising $N$.

\subsection{Fuzzy surfaces with higher order genus}

With the fuzzy vertex, higher order genus surfaces can be concatenated:

Firstly, the functions $f_{m}(q)$ can be concatenated with mirrored
functions $f_{m}(q_{E}-q)$ to generate a surface, in which a cylinder
splits into two cylinders that surrounds a hole and again merge into
one cylinder. The $\hat{X}$ and $\hat{Y}$ matrices of this fuzzy
one-hole space would have an upper part looking like the matrices
shown in Fig. 4 and a lower part, which is mirrored at the $n+m=0$
axis of these matrices.

As described above, such a one-hole surface can be deformed at the
ends by letting the functions $f_{m}(q)$ go to $0$ at the ends to
form a (deformed) torus. In this way we have found a deformed fuzzy
torus with classical limit for $N\rightarrow\infty$. 

For surfaces with genus $g$, $g$ one-hole surfaces can be concatenated
to a surface with $g$ holes and the ends of the result can be closed.
Thus, fuzzy spaces with a classical limit having an arbitrary genus
can be constructed. A similar result was obtained in \cite{Arnlind:2009}
in a more algebraic way.

\section{Comparison with known fuzzy spaces}

After presenting one of the main results in the last chapter, we compare
the new type of fuzzy spaces with already known fuzzy spaces. In the
next two sections, we transform a fuzzy cylinder and a fuzzy torus
by a coordinate transformation into a form, in which we expect to
find fuzzy string vertices within the resulting matrices. We will
see that these matrices are composed of parts, which are rather similar
to the ones constructed in the previous chapter.

\subsection{Coordinate transformed fuzzy cylinder}

We start with an ordinary, classical cylinder aligned along the $z$-axis
and parametrized by $x(z,\varphi)=\cos\varphi$ and $x(z,\varphi)=\cos\varphi$.
A coordinate transformation for deforming the cylinder into a U-shaped
surface is
\begin{equation}
z'=\alpha z^{2}-x\ y'=y\ x'=z\label{eq:cyl_coord_trafo}
\end{equation}
Fig. 5 shows this surface for $\alpha=\frac{1}{3}$ and it can be
seen that cross-sections orthogonal to the $z$-axis (from the bottom
up) are a circle transitioning into a $8$, which goes over into two
circles. A string vertex is part of the deformed cylinder. One would
expect that a correspondingly transformed fuzzy cylinder shows an
analogous behavior.

\begin{figure}

\includegraphics[scale=0.5]{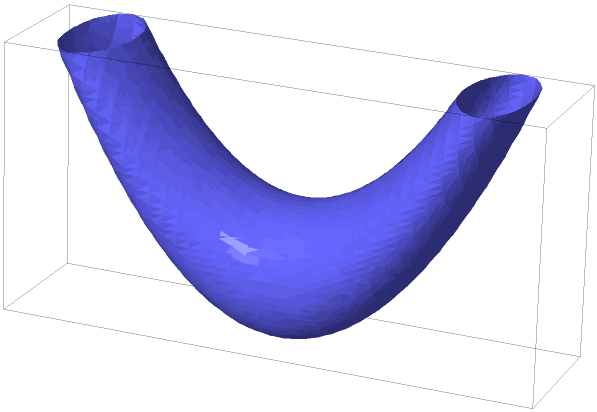}\caption{U-shaped surface of a coordinate transformed cylinder}
\end{figure}
A corresponding fuzzy cylinder can be defined by the Hermitian $N\times N$
matrices $\hat{X}$, $\hat{Y}$ and $\hat{\text{Z}}$ with entries
\begin{equation}
\hat{Z}=\sum_{n=1}^{N}5\left(-1+\frac{2n}{N}\right)\hat{e}_{nn}\ \hat{X}=\sum_{n=1}^{N}\frac{1}{2}\left(\hat{e}_{n,n+1}+\hat{e}_{n+1,n}\right)\ \hat{Y}=\sum_{n=1}^{N}\frac{i}{2}\left(\hat{e}_{n,n+1}-\hat{e}_{n+1,n}\right)
\end{equation}
see (\ref{eq:def_cylinder}) or \cite{Steinacker:2011}, for example.
As mentioned several times, the fuzzy cylinder as defined here converges
only within border. However in the following, we are only interested
in the middle part of the cylinder, which will be deformed to a ``U''.

We apply now the same coordinate transformation (\ref{eq:cyl_coord_trafo})
to the fuzzy cylinder 
\begin{equation}
\hat{Z}'=\alpha\hat{Z}^{2}-\hat{X}
\end{equation}
and diagonalize the matrix $\hat{Z}$ with a unitary matrix $\hat{P}$.
With this we define
\begin{equation}
\hat{Y}'=\hat{P}^{\dagger}\hat{Y}\hat{P}\;\;\;\hat{X}'=\hat{P}^{\dagger}\hat{Z}\hat{P}\label{eq:unitary_transform_fuzzy_cyl}
\end{equation}

We have performed this transformations numerically for $N=40$ and
$\alpha=5/8$. A solver was used, which calculated real eigenvectors
for the eigenvalues of $\hat{Z}'$. In such a way, a unitary transformation
could be assembled that guaranted that the matrix entries of $\hat{X}'$
are real and the ones of $\hat{Y}'$ are imaginary. We furthermore
ordered the resulting matrices according to increasing eigenvalues
of $\hat{Z}'$. 

\begin{figure}
\includegraphics[scale=0.3]{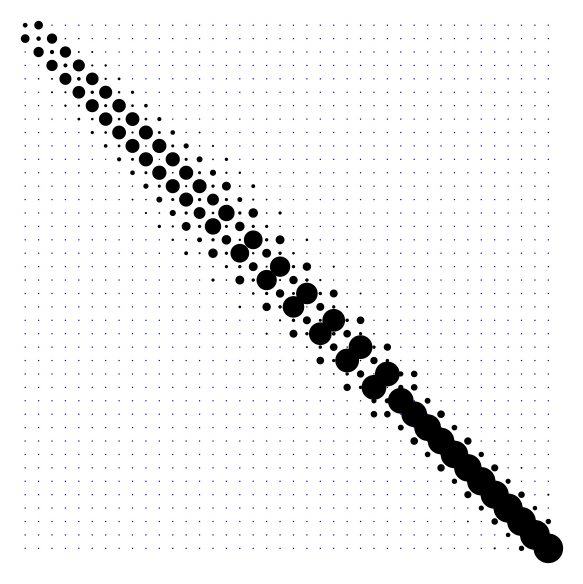}\includegraphics[scale=0.3]{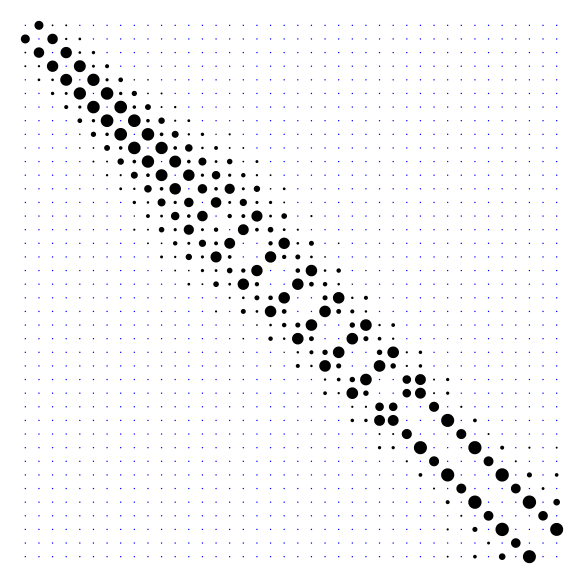}\includegraphics[scale=0.3]{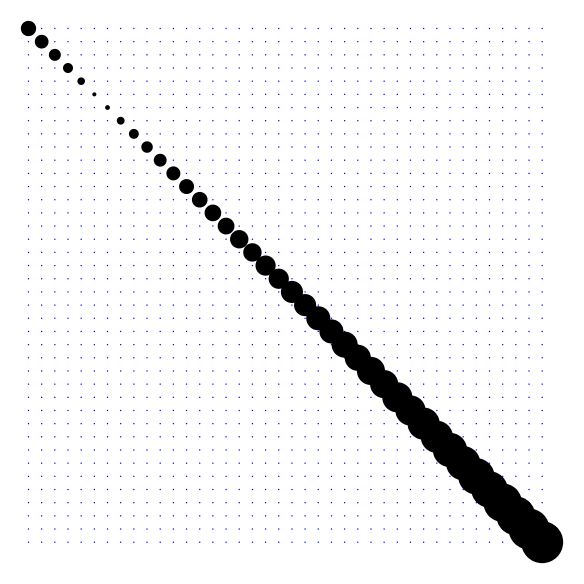}
\caption{Visualization for coordinate transformed fuzzy cylinder}
\end{figure}

Fig. 6 shows the result with dot matrix diagrams for $\hat{X}'$,$\hat{Y}'$
and $\hat{Z}'$. The $2\times2$-matrix structure is visible very
well. Comparing the diagrams with Fig. 2, 3 and 4, it is clear that
they represent a fuzzy space, which has Fig. 5 as classical limit.
In the lower right part, where the $\hat{Z}$-values become ambiguous
(i.e. the part of the double cylinder), the transformation $\hat{P}$
is not any more unique. A block transformation (\ref{eq:unitary_block})
from the two interlaced cylinders to two ordinary cylinders had been
produced by the solver accidently.

In summary, the coordinate transformed fuzzy cylinder can be seen
as a regularization of a matrix-valued functions.

\subsection{Projected fuzzy Clifford torus}

Another very well known space is the fuzzy Clifford torus \cite{Shimada:2003,Steinacker:2011}.
It can be defined with 4 Hermitian $N\times N$ matrices

\begin{equation}
\hat{X}^{1}=\sum_{n=1}^{N}\frac{a}{2}\left(\hat{e}_{n,n+1}+\hat{e}_{n+1,n}\right)\ \ \ \hat{Y}^{1}=\sum_{n=1}^{N}\frac{ia}{2}\left(\hat{e}_{n,n+1}-\hat{e}_{n+1,n}\right)
\end{equation}
\begin{equation}
\hat{X}^{2}=\sum_{n=1}^{N}b\cos\left(\frac{2\pi n}{N}\right)\hat{e}_{nn}\ \ \ \hat{Y}^{2}=\sum_{n=1}^{N}=b\sin\left(\frac{2\pi n}{N}\right)\hat{e}_{nn}
\end{equation}
which can be seen as a regularization of the Clifford torus embedded
in a 4 dimensional space. As in the case of the fuzzy cylinder, we
apply a coordinate transformation to the Clifford torus to project
it into 3 dimensional space. The matrix analogue of the coordinate
transformation is

\begin{equation}
\hat{X}^{I}=\sum_{n=1}^{N}\frac{1}{1+X_{nn}^{4}}\hat{e}_{nn}
\end{equation}

\begin{equation}
\hat{Y}=\hat{X}^{I}\hat{X}^{2}\ \hat{X}=\hat{X}^{I}\hat{X}^{3}\ \hat{Z}=\hat{X}^{I}\hat{X}^{1}
\end{equation}
Again, the matrix $\hat{Z}$ is then diagonalized and $\hat{X}$ and
$\hat{Y}$ are transformed as in (\ref{eq:unitary_transform_fuzzy_cyl})
to arrive at the projected fuzzy Clifford torus.

We have numerically performed this transformations with $a=1$ and
$b=2$. Fig. 7 shows the result for the matrices $\hat{X}$, $\hat{Y}$
and $\hat{Z}$ for $N=40$. A $2\times2$ matrix structure is again
visible and by comparing the diagrams with the previous ones, one
would expect that the fuzzy space is composed of two caps that are
connected with each other via two string vertices, which is the case
for an ordinary torus.

\begin{figure}
\includegraphics[scale=0.3]{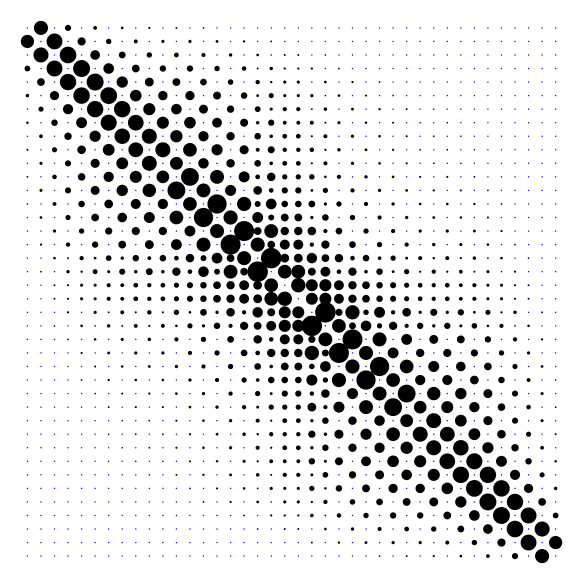}\includegraphics[scale=0.3]{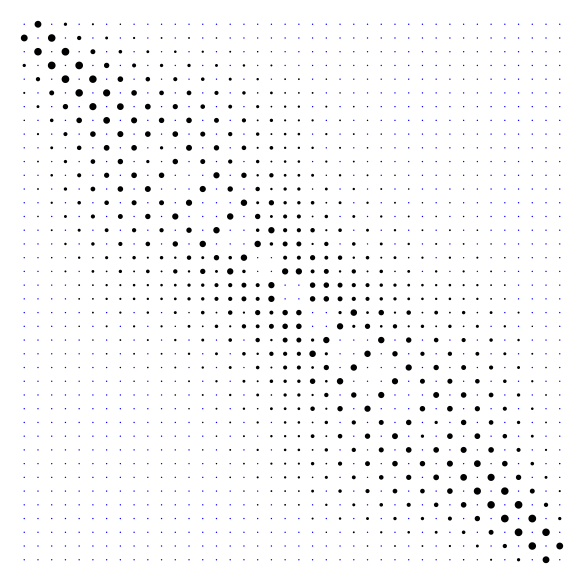}\includegraphics[scale=0.3]{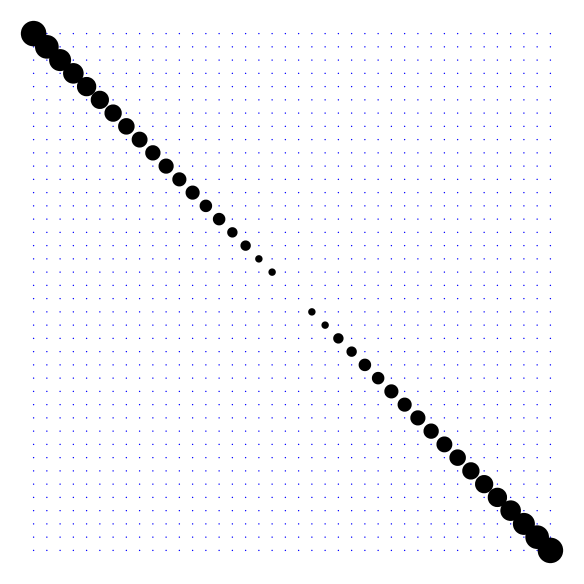}
\caption{Visualization for coordinate transformed fuzzy torus}
\end{figure}

\subsection{Graph-based fuzzy vertex}

In the end we will use the formalism developed so far for finding
a classical limit for the fuzzy spaces defined in \cite{Sykora:2016},
which was actually the main motivation for this work. The building
block of these fuzzy spaces was the so-called fuzzy vertex with matrices
$\hat{X}$ and $\hat{Y}$, which are both of the form 

\begin{equation}
\hat{F}=\left(\begin{array}{cccc|ccccccc}
\ddots & \ddots &  & \\
\ddots & 0 & r^{A} & \\
 & \overline{r^{A}} & 0 & r^{A}\\
 &  & \overline{r^{A}} & 0 & r & r\\
\hline  &  &  & \overline{r} & -x^{B} &  & r^{B}\\
 &  &  & \overline{r} &  & x^{B} &  & r^{B}\\
 &  &  &  & \overline{r^{B}} &  & -x^{B} &  & r^{B}\\
 &  &  &  &  & \overline{r^{B}} &  & x^{B} &  & r^{B}\\
 &  &  &  &  &  & \overline{r^{B}} &  & -x^{B} &  & \ddots\\
 &  &  &  &  &  &  & \overline{r^{B}} &  & \ddots\\
 &  &  &  &  &  &  &  & \ddots
\end{array}\right)
\end{equation}
while the matrix $\hat{Z}$ is diagonal and is composed of $2\times2$
blocks being a multiple of the identity in the lower right part (which
part is indicated by the lines in $\hat{F}$). The dots indicate that
there can be a number of equal entries. However, the matrices $\hat{X}$,
$\hat{Y}$ and $\hat{Z}$ are finite dimensional. After a unitary
$2\times2$ block transformation (\ref{eq:block_transformation})
with (\ref{eq:interlacing_unitary}) for the lower right part, which
leaves the diagonal matrix $\hat{Z}$ unchanged, the result is

\begin{equation}
\hat{F}'=\left(\begin{array}{cccc|ccccccc}
 & \ddots &  & \\
\ddots &  & r^{A} & \\
 & \overline{r^{A}} &  & r^{A}\\
 &  & \overline{r^{A}} &  & r'\\
\hline  &  &  & \overline{r}' &  & x^{B} &  & r^{B}\\
 &  &  &  & x^{B} &  & r^{B}\\
 &  &  &  &  & r^{B} &  & x^{B} &  & r^{B}\\
 &  &  &  & r^{B} &  & x^{B} &  & r^{B} &  & \ddots\\
 &  &  &  &  &  &  & r^{B} &  & x^{B}\\
 &  &  &  &  &  & r^{B} &  & x^{B} &  & \ddots\\
 &  &  &  &  &  &  & \ddots &  & \ddots
\end{array}\right)
\end{equation}
with $r'=\sqrt{2}r$. When we set $r'=r^{A}$, the matrix $\hat{F}'$
is of the form (\ref{eq:matrix_reg}) and can be seen as matrix regularization
of a $2\times2$ matrix-valued function with suitable functions. The
upper left part, which is a fuzzy cylinder, and the lower right part,
which are two interlaced fuzzy cylinders, also can be interpolated
with each other analogously to section \ref{sub:A-fuzzy-vertex}.
We can choose functions $\theta_{1}$, $\theta_{2}$ and $\alpha$
in such a way that they change only for the $q$-values between the
matrix entry of $r'$ and the matrix entry of the first $x^{B}$ of
the right lower part of $\hat{F}'$. With such functions, after regularization
with the same $N$ the matrix $\hat{F}'$ is reproduced. On the other
hand with growing $N$, there exists a series of matrices of arbitrary
size, which has a string vertex as classical limit. Thus, each fuzzy
space defined in \cite{Sykora:2016} is a member of a series of matrix
algebras with classical limit. This classical limit has the same topology
as the corresponding zero-modes surface.

\paragraph*{Acknowledgments and remarks}

The author thanks Harold Steinacker for useful discussions and for
pointing out the relevance of the problem of finding a classical limit
for the fuzzy spaces of \cite{Sykora:2016} , which is solved in the
last section of this article.

Everyone who is interested in the computer programs that have been
used for calculating the matrices and their visualizations is invited
to contact me. I will then send him the SageMath \cite{SageMath}
notebooks. 

Finally, I will thank my little daughter for bringing my back to earth.
One Sunday breakfast we talked about the calculations I do and I started
to try to explain her the basics of topology with the standard examples
of cups with handles and donuts. Then I tried to explain her the creation
of fuzzy surfaces of higher genus by gluing of donuts. She then said:
``I understand, but why should one glue donuts?''


\begin{thebibliography}{10}
\bibitem{Doplicher:1995}S. Doplicher, K. Fredenhagen, J. Roberts
, \textit{The Quantum structure of space-time at the Planck scale
and quantum fields}, Commun. Math. Phys. 172 (1995) {[}hep-th/0303037{]}

\bibitem{Connes:1994}A. Connes, \textit{Non-commutative geometry},
Boston, MA: Academic Press, 1994

\bibitem{Madore:1999}J. Madore, \textit{An Introduction to Noncommutative
Differential Geometry and its Physical Applications}, 2nd edition,
Cambridge University Press, 1999

\bibitem{Ishibashi:1997} N. Ishibashi, H. Kawai, Y. Kitazawa, and
A. Tsuchiya, \textit{A Large N reduced model as super- string}, Nucl.
Phys. B498 , 467 (1997), {[}hep-th/9612115{]}

\bibitem{Banks:1997} T. Banks, W. Fischler, S. H. Shenker, and L.
Susskind, \textit{M theory as a matrix model: A Conjecture}, Phys.
Rev. D55 , 5112 (1997), {[}hep-th/9610043{]}

\bibitem{Steinacker:2011}H. Steinacker, \textit{Non-commutative geometry
and matrix models}, PoS QGQGS \textbf{2011} (2011) 004 {[}arXiv:1109.5521
{[}hep-th{]}{]}

\bibitem{Madore:1991}J. Madore, \textit{The Fuzzy sphere}, 1991,
Class.Quant.Grav. 9 (1992) 69-88

\bibitem{Arnlind:2012}J. Arnlind, J. Hoppe, G. Huisken, \textit{Multi-linear
formulation of differential geometry and matrix regularizations} J.Diff.Geom.
91 (2012) no.1, 1-39 {[}arXiv:1009.4779 {[}math.DG{]}{]}

\bibitem{Arnlind:2009}J. Arnlind, M. Bordemann, L. Hofer, J. Hoppe
and H. Shimada, \textit{Fuzzy Riemann surfaces}, JHEP \textbf{0906}
(2009) 047 {[}hep-th/0602290{]}

\bibitem{Shimada:2003}H. Shimada, \textit{Membrane topology and matrix
regularization}, Nucl. Phys. B \textbf{685} (2004) 297 {[}hep-th/0307058{]}

\bibitem{Sykora:2016}A. Sykora, \textit{The fuzzy space construction
kit}, Oct 3, 2016. 26 pp. {[}arXiv:1610.01504 {[}hep-th{]}{]} 

\bibitem{SageMath}\textit{SageMath, the Sage Mathematics Software
System (Version 7.0)}, 2016, {[}http://www.sagemath.org{]}\end{thebibliography}
\end{document}